\documentclass[aps,prd,preprintnumbers,nofootinbib,11pt]{revtex4-1}

\usepackage{subfig}
\usepackage[export]{adjustbox} 

\usepackage{amsmath,amsfonts,amssymb}  
\usepackage{bm}  
\usepackage{latexsym}  
\usepackage{dcolumn}  
\usepackage{graphicx,epsfig}  
\usepackage{slashed}  
\usepackage{url,hyperref}  
\usepackage{mathrsfs}  

\usepackage{tabularx}  
\usepackage{multirow}  
\usepackage{booktabs}  
\usepackage{threeparttable}  
\usepackage{float}  
\usepackage{subfig}  

\usepackage{graphicx}
\usepackage{caption}
\usepackage{float}
\usepackage{amsmath}
\usepackage{graphicx}
\usepackage{subfig}
\usepackage{float} 
\usepackage{flafter}  

\usepackage{titlesec}
\usepackage{xstring}

\newcommand{\be}{\begin{eqnarray}}
\newcommand{\ee}{\end{eqnarray}}
\newcommand{\bea}{\begin{eqnarray}}
\newcommand{\eea}{\end{eqnarray}}

\usepackage{color}
\definecolor{darkred}{rgb}{.8,0,0}

\definecolor{darkblue}{rgb}{0,0,.7}

\definecolor{darkgreen}{rgb}{0,.7,0}

\usepackage{titlesec}
\begin{document}

%
%
\title{Linking interior curvature to observable shadows: A case study of nonsingular black holes}
%
%
%
%
%
\author{{Ming-Xin Li}$^{1}$}\email[email:~]{lmx842275094@163.com}
\author{Jin Pu$^{1}$}\email[email:~]{pujin@cwnu.edu.cn}
\author{Yi Ling$^{2,3}$} \email[email:~]{lingy@ihep.ac.cn}
\author{Guo-Ping Li$^{1}$} \email[email:~]{gpliphys@yeah.net}
\affiliation{$^1$School of Physics and Astronomy, China West Normal University, Nanchong 637002, China\\
$^2$Institute of High Energy Physics, Chinese Academy of Sciences, Beijing 100049, China\\
$^3$School of Physics, University of Chinese Academy of Sciences, Beijing 100049, China}

\begin{abstract}
We establish a direct connection between the interior curvature structure of nonsingular black holes (BHs) with a Minkowski core and their observable optical signatures. By classifying these spacetimes into three fundamental types, \textbf{Type~I} (Kretschmann scalar $K_{\text{max}}$ increasing with mass $M$), \textbf{Type~II} (mass-independent $K_{\text{max}}$), and \textbf{Type~III} ($K_{\text{max}}$ decreasing with $M$), we demonstrate how subtle variations in the core geometry imprint distinguishable features on the BH shadow. A detailed analysis of photon dynamics reveals that the parameters $\alpha$ and $n$, which control the deviation from Schwarzschild geometry and the radial decay of the regularizing factor, respectively, systematically alter the properties of the photon sphere. These intrinsic geometric differences propagate outward: for fixed parameters, \textbf{Type~III} BHs, with the most compact photon sphere, produce the smallest and brightest shadows, whereas \textbf{Type~I} BHs yield the largest and dimmest ones. Shadow computations under both static and infalling spherical accretion models confirm that the curvature-based classification directly corresponds to observable differences. Critically, \textbf{Type~III} BHs exhibit the strongest sensitivity to parameter variations, making them optimal probes for constraining the underlying spacetime geometry. Our work reveals that even among nonsingular BHs sharing the same asymptotic core, differences in internal curvature are reflected in the shadow morphology, thereby providing a new pathway to test quantum-gravity-inspired models using upcoming high-resolution observations.
\end{abstract}
\maketitle

\newpage  
%
\section{Introduction}
\label{sec1}

Since the advent of General Relativity (GR), BHs have served as fundamental probes of strong-field gravity. The direct detection of BH mergers by LIGO/Virgo \cite{1,2} and the imaging of BH shadows at M87* and Sgr A* by the Event Horizon Telescope (EHT) \cite{3,4,5,6} have transformed BHs from theoretical constructs into observable astrophysical objects. These advances not only confirm GR's predictions in strong-gravity regimes but also provide precise constraints on BH parameters, such as mass, spin, and charge—deepening our understanding of compact systems.

The shadow of a BH, a dark region encircled by lensed photon orbits, has become a powerful geometric diagnostic. Early work by Synge \cite{7} derived the photon capture radius for a Schwarzschild BH, and Luminet \cite{8} performed the first numerical simulations of its optical appearance. Bardeen's extension to Kerr spacetime \cite{9} showed how frame-dragging distorts the shadow, stimulating extensive studies of shadows in various BH solutions \cite{10,11,12,13,14,15,16,17,18}. Subsequent research has expanded the framework to higher dimensions and modified-gravity theories, including high-dimensional spacetimes \cite{19,20}, conformal gravity \cite{21}, Chern-Simons type models \cite{22,23,24}, Einstein-Maxwell-scalar theories \cite{25}, Rastall gravity \cite{26,27}, and Gauss-Bonnet gravity \cite{28,29}. Investigations of exotic compact objects, such as naked singularities \cite{30,31,32,33,34} and wormholes \cite{35,36,37,38,39,40,41,42,43} further illustrate how shadow morphology encodes deviations from classical BH geometry, establishing shadow analysis as a key tool for testing gravitational theories and cosmic censorship.

A central theoretical issue remains the presence of spacetime singularities inside BHs, where curvature divergences challenge both GR and quantum mechanics. Hawking radiation \cite{44}, while confirming the quantum nature of BHs, intensifies this tension through the information loss paradox \cite{45,46,47,48,49,50,51}. Although a complete theory of quantum gravity is still lacking, nonsingular BH models offer interim resolutions via two main approaches: ($\romannumeral1$) semiclassical regularization, which introduces nonsingular energy momentum sources into Einstein's equations \cite{52,53,54,55,56,57,58}, and ($\romannumeral2$) explicit quantum gravity corrections, which modify the classical metric through loop quantum gravity or asymptotic safety methods \cite{59,60,61,62,63,64}. These two classes differ in phenomenology: semiclassical models preserve GR-like dynamics on macroscopic scales, whereas quantum corrected metrics exhibit deviations already at the Planck scale. Among these, nonsingular BHs with a Minkowski core are of particular interest as they represent scenarios where quantum effects render the core asymptotically flat. However, even within this specific class, the detailed behavior of curvature invariants like the Kretschmann scalar can vary significantly, suggesting a potential sub-classification based on interior geometry.

While previous studies have distinguished between singular and nonsingular BHs \cite{65,66,67,68,69,70,71,72,73,74}, or between different asymptotic core types (Minkowski vs. de Sitter) \cite{75,76}, a systematic investigation of how variations within the same core type affect observables has been lacking. Specifically, it remains unclear whether nonsingular BHs sharing the same Minkowski core but differing in their detailed curvature profiles can be observationally distinguished. This work addresses this gap by introducing a curvature-invariant classification scheme based on the mass-scaling of the Kretschmann scalar $K_{\text{max}}$, and systematically mapping these interior geometric types to their observable shadow signatures under both static and infalling spherical accretion models.

The paper is organized as follows. Sec.\ref{sec2} classifies nonsingular BHs with a Minkowski core into three types based on their interior curvature and examines their spacetime properties. Sec.\ref{sec3} studies the optical characteristics of these BHs in spherical accretion models. Conclusions and outlook are presented in Sec.\ref{sec4}.

\section{Spacetime geometry and curvature classification}
\label{sec2}
The modern paradigm of nonsingular BHs traces back to Bardeen's seminal work \cite{77}, which replaced the singular mass parameter in the Schwarzschild solution with a nonsingular mass function $m(r)$. This foundational approach was later placed on a firm field-theoretic foundation by Ay\'on-Beato and Garc\'\i{}a \cite{78}, who demonstrated that Bardeen's metric can be derived from a magnetic monopole in nonlinear electrodynamics. Subsequent generalizations have produced two broad families of nonsingular interiors: ($\romannumeral1$) de Sitter cores, where $m(r)\propto r^3$ as $r\to 0$, yielding a finite energy density $\rho_0$ and pressure $P_0 = -\rho_0$ \cite{79}; and ($\romannumeral2$) Minkowski cores, described by exponentially suppressed mass functions of the type $ m(r) \sim  \mathrm{e}^{-\frac{\alpha}{r} }$, which yield an asymptotically flat interior with vanishing energy density at the origin \cite{79}. In this work we focus on the latter class.

\subsection{Metric and curvature-based classification}
\label{sec2-1}

We consider a specific class of nonsingular BHs that share a Minkowski core but differ in their detailed curvature structure, as captured by the Kretschmann scalar. For a static, spherically symmetric spacetime the line element can be expressed as
\begin{equation}
\mathrm{d}s^{2} = -f(r)\mathrm{d}t^{2} + f(r)^{-1}\mathrm{d}r^{2} + r^{2}(\mathrm{d}\theta^{2} + \sin^{2}\theta\,\mathrm{d}\phi^{2}),
\label{eq:1}
\end{equation}
where the metric function $f(r)$ is chosen so that $f(r) \to 1$ as $r \to 0$, ensuring a nonsingular Minkowski core. Early inspiration for such models came from the Generalized Uncertainty Principle (GUP), which suggests significant quantum gravity modifications near the Planck scale \cite{48}. A representative metric function takes the form \cite{48}
\begin{equation}
f(r)=1 - \frac{2M \, \mathrm{e}^{-\frac{\alpha}{r^{2}}}}{r},
\label{eq:2}
\end{equation}
with ADM mass $M$ and a dimensionless parameter $\alpha $ ($\alpha=0$ recovers the Schwarzschild metric). The corresponding mass function $m(r)=M \mathrm{e}^{ -\frac{\alpha}{r^2} }$ exhibits exponential suppression at small $r$. Other models propose different suppressions, e.g. $m(r)=M \mathrm{e}^{-\frac{\alpha}{r} }$ or $m(r)=M \mathrm{e}^{-\frac{\alpha M^{2/3}}{r^{2}} }$ \cite{79,80}, all satisfying $ m(r) \to 0 $ as $ r \to 0 $ and hence a vanishing energy density $ \rho(r) = m'(r)/(4\pi r^2) $ at the core. A unified parametrisation that encompasses these cases is \cite{80}
\begin{equation}
f(r) =1+2\psi (r) =1 - \frac{2M \mathrm{e}^{-\frac{\alpha M^{\beta}}{r^{n}}}}{r},
\label{eq:3}
\end{equation}
with dimensionless parameters obeying $n> \beta \ge  0$ and $n \ge 1$. The parameter $n$ governs the fall-off rate of the exponential regulator in the mass function $m(r)$; a larger $n$ leads to a steeper suppression of mass near the origin $r \to 0$.

Horizons correspond to the roots of $f(r_{h})=0$. Solving this yields the inner and outer horizon radii as explicit functions of the mass,
\begin{equation}
r_{h\pm} = 2M \left[\frac{\theta}{W_{k}(\theta)} \right]^{\frac{1}{n}}, \qquad\theta = - \frac{\alpha n M^{\beta -n}}{2^n}, \qquad (k=0,-1)
\label{eq:4}
\end{equation}
where $W_{k}$ denotes the $k-$th branch of the Lambert-$W$ function. For negative arguments ($\theta <0$) two real branches exist: $k=-1$ (inner horizon, $W_{-1}(\theta)\le -1$) and $k=0$ (outer horizon, $W_{0}(\theta)\ge -1$) \cite{74,79}. Because shadow properties are dominated by the outer horizon, we restrict attention to the $k=0$ branch. For this branch the series representation
\begin{equation}
W_{0}(\theta) = \sum_{n=1}^{\infty } \frac{(-n)^{n-1}}{n!}\theta ^n,
\label{eq:5}
\end{equation}
with $W_{0}(\theta)\ge -1$ \cite{81}, is useful. The existence of real horizons imposes physical constraints on the parameter space. From the condition for the merger of the inner and outer horizons ($ r_{h-}=r_{h+}$), which occurs at the extremal limit of the Lambert-$W$ function when its argument reaches the critical value $\theta = - \mathrm{e}^{-1}$ \cite{48}, we obtain a minimum mass threshold
\begin{equation}
M \ge \left( \frac{\alpha n \mathrm{e}}{2^n} \right)^{\frac{1}{n-\beta}}.
\label{eq:6}
\end{equation}
Saturation of this bound ($M=M_{min}$) marks the extremal limit, where the inner and outer horizons coincide. From the extremal condition one obtains the allowed range of $\alpha$,
\begin{equation}
0 \le \alpha \le \frac{2^n M^{n-\beta}}{n \mathrm{e}}.
\label{eq:7}
\end{equation}

\begin{figure}[H]
\centering
\vspace{-5pt}
    \includegraphics[width=0.3\linewidth]{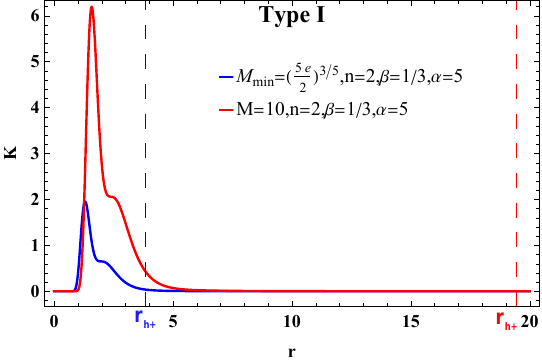}
    \label{Fig1a} 
    \includegraphics[width=0.305\linewidth]{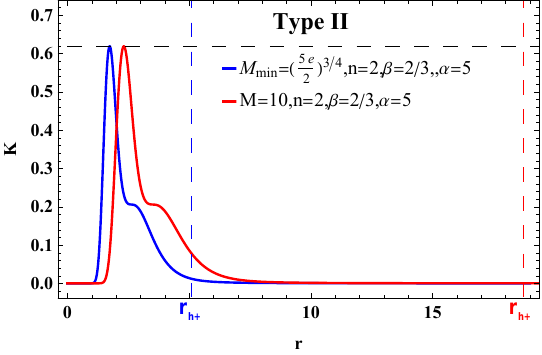}
    \label{Fig1b}
    \includegraphics[width=0.31\linewidth]{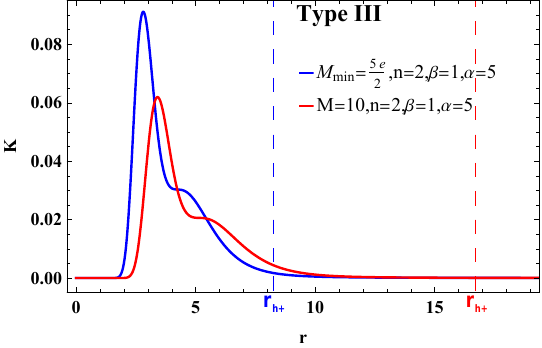}
    \label{Fig1c}
    \vspace{-20pt}
\caption{ The Kretschmann scalar $K$ as a function of $r$ for the three types of nonsingular BHs with a Minkowski core. }
 \vspace{-20pt}
\label{Fig:1}
\end{figure}

To quantify differences in internal spacetime structure we examine the Kretschmann scalar $K=R_{\mu\nu\rho\lambda }R^{\mu\nu\rho\lambda }$, a curvature invariant that remains finite as $r \to 0$. For the metric (\ref{eq:3}) it reads
\begin{align}
K
&= 4 M^{2} r^{-6} \mathrm{e}^{-2 \alpha M^{\beta} r^{-n}}
\biggl[
\alpha^{4} n^{4} M^{4\beta} r^{-4n}
- 2 \alpha^{3} n^{3} (n + 3) M^{3\beta} r^{-3n}  \notag \\
&\quad
+ \alpha^{2} n^{2} (n^{2} + 6n + 17) M^{2\beta} r^{-2n}
- 4 \alpha n (n + 5) M^{\beta} r^{-n}
+ 12
\biggr].
\label{eq:8}
\end{align}
As shown in Ref.~\cite{80}, the maximum of the Kretschmann scalar scales as $K_{\text{max}} \propto M^{(2 - \frac{6\beta}{n})}\alpha^{-\frac{6}{n}}$. This scaling implies a distinct functional dependence of $K_{\text{max}}$ on the BH mass $M$, which leads to a natural three-fold classification of nonsingular BHs with a Minkowski core (illustrated in Fig.~\ref{Fig:1}):
\begin{itemize}
 \vspace{-10pt}
    \item[($\romannumeral1$)] \textbf{Type I} ($0 \le \beta < n/3$): $K_{\text{max}}$ increases with $M$;
     \vspace{-10pt}
    \item[($\romannumeral2$)] \textbf{Type II} ($\beta = n/3$): $K_{\text{max}}$ is independent of $M$;
     \vspace{-10pt}
    \item[($\romannumeral3$)] \textbf{Type III} ($n/3 < \beta < n$): $K_{\text{max}}$ decreases as $M$ increases.
         \vspace{-10pt}
\end{itemize}
This classification reveals fundamental differences in the internal curvature structure. Notably, several models studied earlier fit into this scheme; for example, the metric~(\ref{eq:2}) from Ref.~\cite{48} corresponds to \textbf{Type I}, while the model analysed in Ref.~\cite{75,76} belongs to \textbf{Type II}.
This curvature-based taxonomy is particularly well-founded because the Kretschmann scalar $K \equiv R_{\mu\nu\rho\sigma}R^{\mu\nu\rho\sigma}$ as our classification invariant is motivated by its mathematical properties and physical clarity within our framework. First, as the full quadratic contraction of the Riemann tensor, it provides a comprehensive measure of curvature by incorporating both local matter contributions (via the Ricci tensor) and non-local tidal components (via the Weyl tensor). In four dimensions, this decomposition reads $R_{abcd}R^{abcd} = C_{abcd}C^{abcd} + 2R_{ab}R^{ab} - \frac{1}{3}R^2$. Consequently, $K$ captures the joint curvature content of the interior, which is essential for distinguishing between different regularisation mechanisms that may affect Ricci and Weyl parts differently.

Second, $K$ offers several decisive technical advantages for our purpose. (i) It is non-negative and has well-defined dimensionality ($\text{length}^{-4}$), making the comparison of its peak value $K_{\text{max}}$ and its scaling with mass $M$ unambiguous. This contrasts with the Ricci scalar $R$, which can vanish or change sign, or with Weyl-only invariants that ignore curvature sourced by the effective stress-energy of the regularised core. (ii) For our parametrised metric (Eq.~\ref{eq:3}), $K_{\text{max}}$ exhibits a clean power-law dependence on the BH parameters: $K_{\text{max}} \propto M^{(2 - \frac{6\beta}{n})}\alpha^{-\frac{6}{n}}$. The mass-scaling exponent $p = 2-6\beta/n$ directly encodes how the regularization efficiency varies with $M$, and it provides the natural basis for our three-fold taxonomy (\textbf{Type I/II/III}). (iii) In regular spacetimes with a Minkowski core, $K$ remains finite everywhere and typically displays a single, pronounced maximum near the core. This makes $K_{\text{max}}$ a robust and sensitive diagnostic of the interior curvature strength, whereas invariants like $R$ might be identically zero at the core for some models, offering no discriminatory power.

\begin{figure}[H]
\centering
\vspace{-15pt}
      \subfloat[\textbf{Type~I}( $\beta = 1/6$)]{\includegraphics[width=0.3\linewidth]{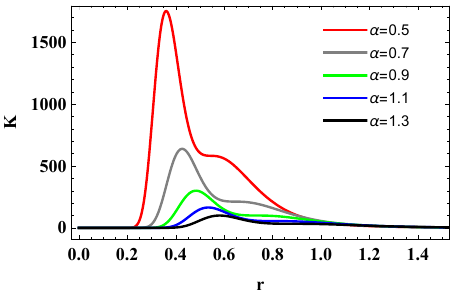}}\hfill
    \subfloat[\textbf{Type~II}($\beta = 2/3$)]{\includegraphics[width=0.3\linewidth]{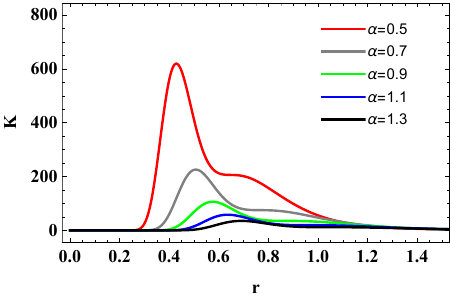}}\hfill
    \subfloat[\textbf{Type~III}($\beta = 3/4$)]{\includegraphics[width=0.3\linewidth]{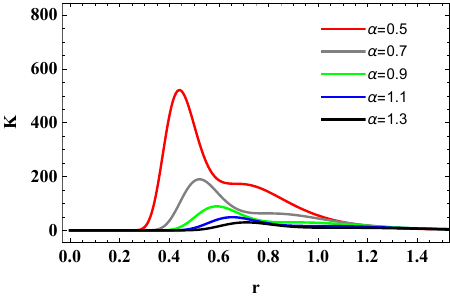} }
    \vspace{-5pt}
\caption{ Dependence of the Kretschmann scalar $K$ on $\alpha$ for the three BH types ($n=2$, $M=2$). }
\vspace{-15pt}
\label{Fig:2}
\end{figure}

Fig.~\ref{Fig:2} displays the influence of the parameter $\alpha$ on the Kretschmann scalar. For a fixed value of $\alpha$, \textbf{Type~I} BHs exhibit the highest curvature and \textbf{Type III} the lowest. All three BH types, however, share a monotonic trend: $K_{\text{max}}$ decreases as $\alpha$ increases. When $\alpha = 0$ the spacetime reduces to Schwarzschild, implying that $\alpha$ can be interpreted as a measure of quantum gravity effects \cite{48,75,76,79,80}. Larger $\alpha$ therefore leads to a smoother geometry near the core, effectively regularizing the central singularity.

\begin{figure}[H]
\centering
\vspace{-15pt}
    \subfloat[\textbf{Type~I}]{%

        \includegraphics[width=0.3\linewidth]{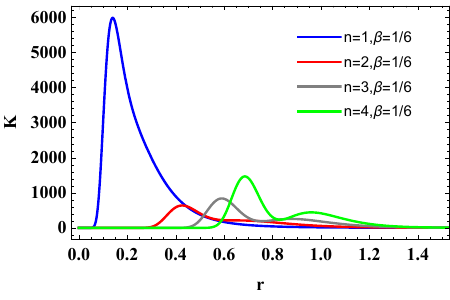}
    }\hfill
    \subfloat[\textbf{Type~II}]{%

        \includegraphics[width=0.3\linewidth]{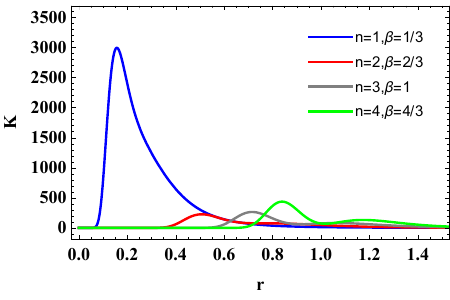}
    }\hfill
    \subfloat[\textbf{Type~III}]{%

        \includegraphics[width=0.3\linewidth]{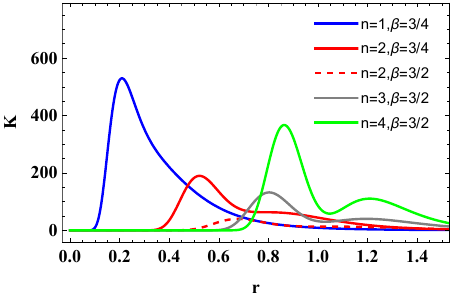}
    }
    \vspace{-6pt}
\caption{Dependence of $K(r)$ on $n$ for the three BH types ($M=2$, $\alpha=0.7$).}
\vspace{-22pt}
\label{Fig:3}
\end{figure}
The dependence of the Kretschmann scalar on $n$, shown in Fig.~\ref{Fig:3}, exhibits more intricate behavior. Note that $\beta$ is tied to $n$ differently for each type: for \textbf{Type~I} we fix $\beta=1/6$; for \textbf{Type~II}, $\beta=n/3$; for \textbf{Type~III} we use $\beta=3/4$ when $n=1,2$ and $\beta=3/2$ when $n=2,3,4$. As $n$ increases from $1$ to $2$, $K_{\text{max}}$ decreases for all types; beyond $n=2$ it rises again. Moreover, the radial position of the curvature maximum shifts outward with increasing $n$. This non-monotonic behaviour, together with the migration of the peak, signals a substantial reorganisation of the spacetime geometry controlled by $n$.

\begin{figure}[H]
\centering
\vspace{-5pt}
    \includegraphics[width=0.3\linewidth]{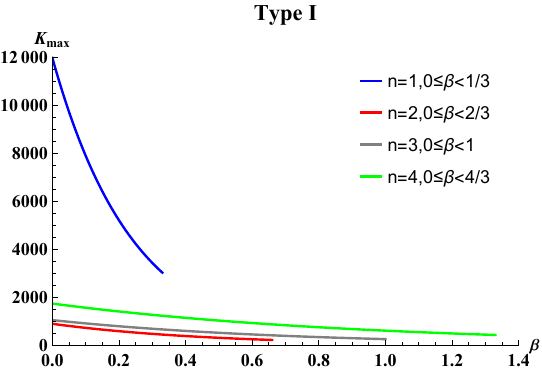}
    \hspace{30pt}
    \includegraphics[width=0.3\linewidth]{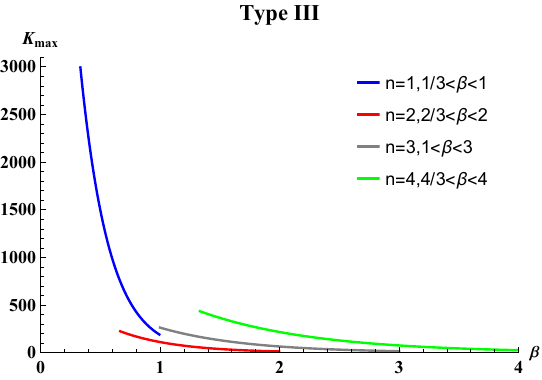}
    \vspace{-5pt}
\caption{ Maximum Kretschmann scalar $K_{\max}$ as a function of $\beta$ for \textbf{Type I} and \textbf{III} BHs ($\alpha=0.7$).}
\vspace{-25pt}
\label{Fig:4}
\end{figure}

Fig.~\ref{Fig:4} shows how $K_{\max}$ varies with $\beta$ for \textbf{Type~I} and \textbf{III} (\textbf{Type~II}, $\beta$ is fixed once $n$ is chosen). With $n$ and $\alpha$ held fixed, $K_{\max}$ decreases as $\beta$ increases, i.e. a larger $\beta$ yields a flatter core. The plots also reproduce the non-monotonic $n$-dependence noted above: $K_{\max}$ drops from $n=1$ to $n=2$, then rises again for $n>2$.

In summary, the parameter $\alpha$ monotonically suppresses the central curvature for all types, with larger $\alpha$ yielding a flatter core geometry. The parameter $n$ exerts a more subtle, non-monotonic influence: curvature is most suppressed near $n\approx2$, while both smaller ($n=1$) and larger ($n>2$) values lead to higher $K_{\max}$. For \textbf{Type~II} the parameter $\beta$ is fixed by $n$; for \textbf{Type~I} and \textbf{III}, a larger $\beta$ also flattens the core. These curvature differences, governed sensitively by $n$ and $\alpha$, are expected to leave imprints on the photon sphere and ultimately on the optical appearance, which we examine next.

\subsection{Photon sphere and effective potential}
\label{sec2-2}

Photon trajectories are governed by null geodesics, which depend on the underlying spacetime curvature. For the metric~(\ref{eq:1}) the equations of motion follow from the Euler-Lagrange equations,
\begin{equation}
\frac{\mathrm{d}}{\mathrm{d}\lambda}\left(\frac{\partial\mathcal{L}}{\partial\dot{x}^\mu}\right) - \frac{\partial\mathcal{L}}{\partial x^\mu} = 0,
    \label{eq:9}
\end{equation}
with affine parameter $\lambda$ and $\dot{x}^{\mu}=dx^{\mu}/d\lambda$; for null geodesics $\mathcal{L}=\frac{1}{2} g_{\mu\nu}\dot{x}^\mu\dot{x}^{\nu}=0$. The spacetime admits two Killing vectors, $\partial_{t}$ and $\partial_{\phi}$, giving conserved energy and angular momentum
\begin{equation}
E=f(r)\left(\frac{\mathrm{d}t}{\mathrm{d}\lambda}\right), \qquad L=r^{2}\left(\frac{\mathrm{d}\phi }{\mathrm{d}\lambda}\right).
    \label{eq:10}
\end{equation}
Confining motion to the equatorial plane ($\theta=\pi/2$, $\dot{\theta}=0$) and rescaling the affine parameter as $\tilde\lambda=\lambda/|L|$ yields the first-order differential equations of motion
\begin{equation}
\frac{\mathrm{d}t}{\mathrm{d}\tilde{\lambda}} = \frac{1}{bf(r)}, \qquad \frac{\mathrm{d}\phi}{\mathrm{d}\tilde{\lambda}} = \pm\frac{1}{r^2}, \qquad \frac{\mathrm{d}r}{\mathrm{d}\tilde{\lambda}} = \sqrt{\frac{1}{b^2} - V_{\text{eff}}(r)}.
\label{eq:11}
\end{equation}
where $b = |L|/E$ is the impact parameter and the effective potential is
\begin{equation}
V_{\text{eff}}(r) =\frac{f(r)}{r^{2} } = \frac{1}{r^{2}} - \frac{2M \, \mathrm{e}^{-\frac{\alpha M^{\beta}}{r^{n}}}}{r^{3}}.
\label{eq:12}
\end{equation}
The photon sphere corresponds to the unstable circular orbit satisfying
\begin{equation}
\begin{aligned}
&\frac{\mathrm{d}r}{\mathrm{d}\tilde\lambda}=0 \qquad &\Rightarrow& \qquad \frac{1}{b_c^2}=V_{\text{eff}}(r_c), \\
&\frac{\mathrm{d}^{2} r}{\mathrm{d}^{2}\tilde{\lambda}}=0 \qquad &\Rightarrow& \qquad \frac{\mathrm{d}V_{\text{eff}}}{\mathrm{d}r}\bigg|_{r=r_c} = 0.
\label{eq:13}
\end{aligned}
\end{equation}
where $r_c$ denotes the photon sphere radius. Substituting Eq. (\ref{eq:12}) into Eq. (\ref{eq:13}), we obtain
\begin{equation}
\mathrm{e}^{\alpha M^{\beta}r_{c}^{-n}} r_{c}^{n+1} +\alpha n M^{\beta+1} =3M r_{c}^{n}.
\label{eq:14}
\end{equation}
The corresponding critical impact parameter, which sets the shadow size for a distant observer, is
\begin{equation}
b_c = \frac{r_{c} }{\sqrt{f(r_{c})} }=r_{c}{(1 - \frac{2M \mathrm{e}^{-\frac{\alpha M^{\beta}}{r_{c}^{n}}}}{r_{c}}) ^{-\frac{1}{2} } }  .
\label{eq:15}
\end{equation}

\begin{figure}[H]
\centering
\vspace{-15pt}
   \subfloat[$n=1$ ]{ \includegraphics[width=0.3\linewidth]{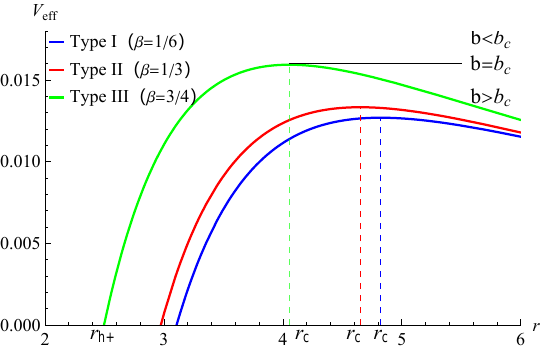}}\hfill
      \subfloat[$n=2 $]{\includegraphics[width=0.3\linewidth]{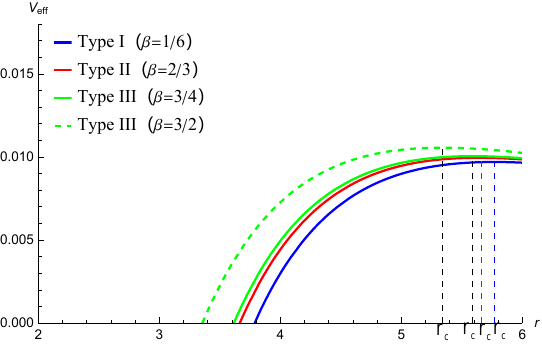}}\hfill
      \subfloat[$n=3 $]{\includegraphics[width=0.3\linewidth]{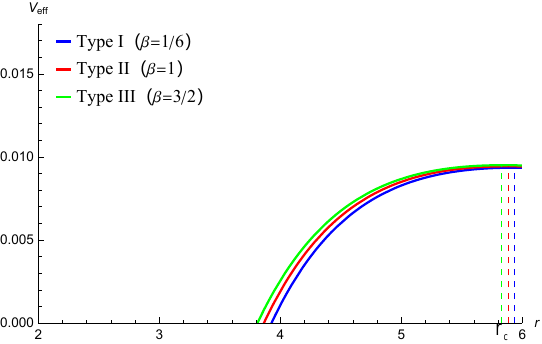}}
      \vspace{-5pt}
 \caption{Effective potential $V_{\text{eff}}(r)$ for three BH types at different values of $n$ ($M=2$, $\alpha=0.7$). }
\vspace{-15pt}
\label{Fig:5}
\end{figure}

Fig.~\ref{Fig:5} illustrates the behavior of the effective potential $V_{\text{eff}}(r)$ for the three types. The potential vanishes at the outer horizon $r_{h+}$, reaches a maximum at the photon sphere radius $r_{c}$, and defines the critical impact parameter $b_{c}$ that marks the shadow boundary. Photons with $b>b_{c}$ are deflected by the potential barrier, those with $b<b_{c}$ cross the horizon and are captured, and photons with $b=b_{c}$ asymptotically approach the unstable circular orbit at $r_{c}$. The peak value of $V_{\text{eff}}(r)$ at $r_c$ is highest for \textbf{Type~III} and lowest for \textbf{Type~I}, reflecting an inverse correlation with the compactness of the photon orbit. Specifically, both $r_{h+}$ and $r_{c}$ are largest for \textbf{Type~I} and most compact for \textbf{Type~III}. Increasing $n$ systematically lowers $V_{\text{eff}}$ while increasing $r_{h+}$ and $r_{c}$ for all three types; concurrently, the differences among the types in $V_{\text{eff}}$, $r_{h+}$ and $r_{c}$ gradually diminish, i.e. the effective potentials and characteristic length scales become more similar for larger $n$.

\begin{figure}[H]
\centering
\vspace{-10pt}
\subfloat[$\alpha=0.3$ ]{ \includegraphics[width=0.3\linewidth]{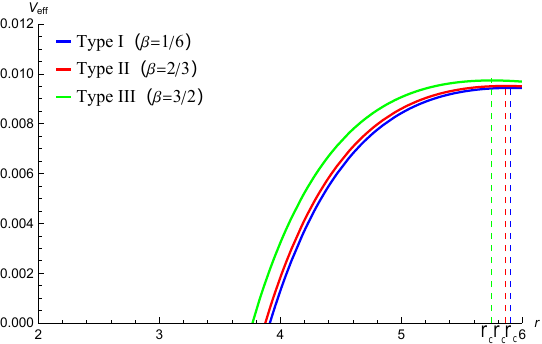}}\hfill
      \subfloat[$\alpha=0.7$]{\includegraphics[width=0.3\linewidth]{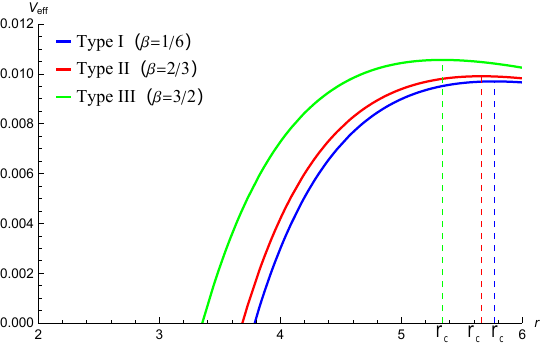}}\hfill
      \subfloat[$\alpha=1$]{\includegraphics[width=0.3\linewidth]{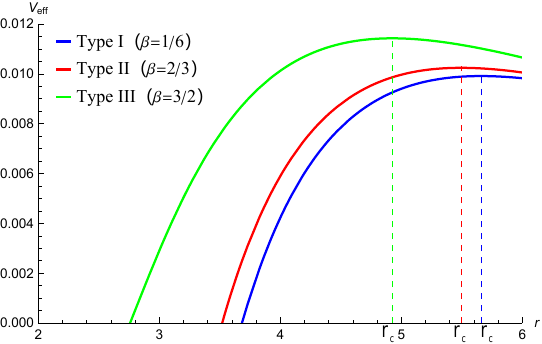}}
\vspace{-5pt}
\caption{The effective potentials $V_{\text{eff}}(r)$ for various values of $\alpha$ ($M=2$, $n=2$).}
\vspace{-10pt}
\label{Fig:6}
\end{figure}

Fig.~\ref{Fig:6} displays the variation of $V_{\text{eff}}(r)$ with $\alpha$. As $\alpha$ increases, $V_{\text{eff}}(r)$ increases, while in contrast, both $r_{h+}$ and $r_c$ decrease. Moreover, the distinctions among the three BH types in terms of $V_{\text{eff}}(r)$, $r_{h+}$ and $r_c$ become progressively larger with increasing $\alpha$, meaning that larger $\alpha$ values lead to more distinct differences in both the effective potential and the characteristic length scales of the three BH types. These systematic changes in $V_{\text{eff}}(r)$, $r_c$, and $b_c$ with $n$ and $\alpha$ establish a geometric basis for the distinct photon orbital structures of the three BH types, which will subsequently manifest in their shadow properties.

\subsection{Orbit classification and critical parameters}
\label{sec2-3}
To understand how the parameters influence photon paths, we classify trajectories according to the total number of orbits $N(b)=\phi/(2\pi)$, where $\phi$ is the total change in azimuthal angle from source to observer \cite{82}. From Eq. (\ref{eq:11}) the orbit equation is
\begin{equation}
\frac{\mathrm{d}r}{\mathrm{d}\phi}
= r^2\sqrt{\frac{1}{b^2}-V_{\text{eff}}(r)}.
\label{eq:16}
\end{equation}
Light rays are categorised as
\begin{itemize}
\vspace{-10pt}
    \item \textbf{Direct emission} (\(N < 3/4\)): the trajectory crosses the equatorial plane only once.
    \vspace{-10pt}
    \item \textbf{Lensed ring} (\(3/4 < N < 5/4\)): the trajectory crosses the equatorial plane twice.
    \vspace{-10pt}
    \item \textbf{Photon ring} (\(N > 5/4\)): the trajectory crosses the equatorial plane at least three times.
        \vspace{-10pt}
\end{itemize}
For numerical integration, it is convenient to set $u = 1/r$, which transforms Eq. (\ref{eq:16}) into
\begin{equation}
\frac{\mathrm{d} u}{\mathrm{d} \phi} = \sqrt{\frac{1}{b^2} - u^2 (1 - 2Mu \mathrm{e}^{-\alpha M^{\beta} u^{n}  })}.
\label{eq:17}
\end{equation}
By integrating the trajectory equation, the azimuthal angle $\phi$ can be expressed as
\begin{equation}
\phi = \int \frac{\mathrm{d}u}{\sqrt{\frac{1}{b^2} - u^2 (1 - 2Mu \mathrm{e}^{-\alpha M^{\beta} u^{n}  })}}.
\label{eq:18}
\end{equation}
\begin{figure}[H]
\centering
\vspace{-12pt}
\subfloat[$n=1$]{\includegraphics[width=0.315\textwidth, valign=t]{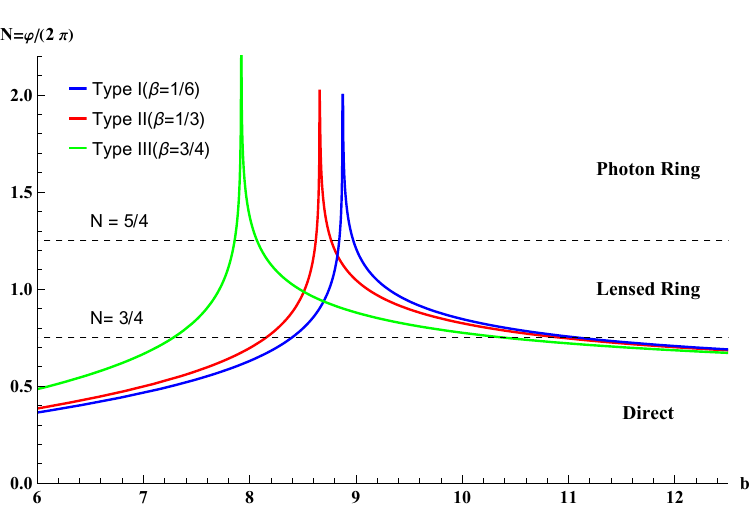}}\hfill
\subfloat[$n=2$]{\includegraphics[width=0.315\textwidth, valign=t]{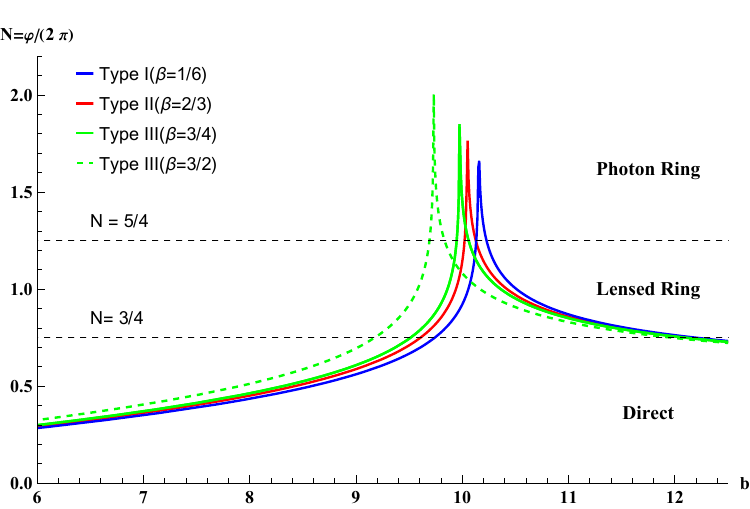}}\hfill
\subfloat[$n=3$]{\includegraphics[width=0.315\textwidth, valign=t]{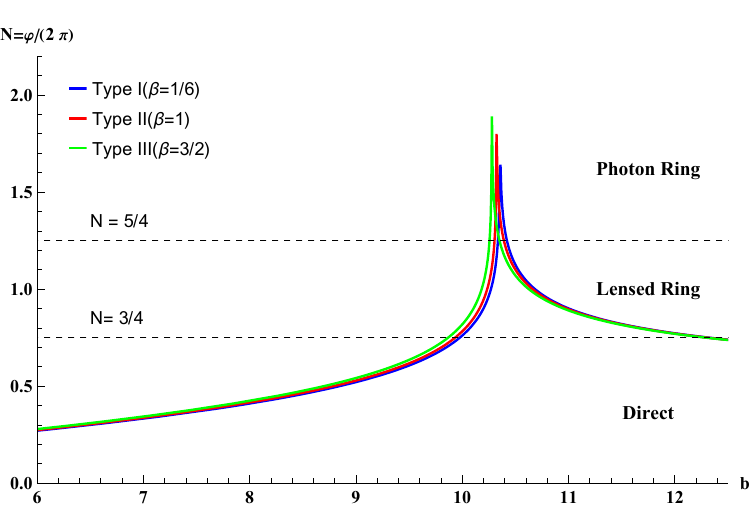}}

\vspace{-13pt}

\subfloat[\textbf{Type I} ($n=1,\beta=1/6$)]{\includegraphics[width=0.315\textwidth, valign=t]{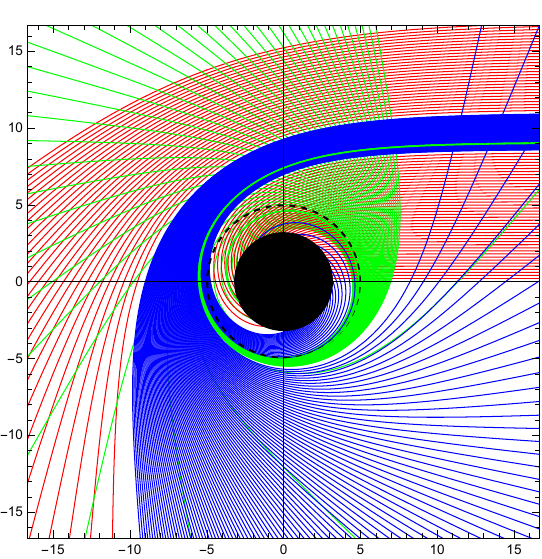}}\hfill
\subfloat[\textbf{Type II} ($n=1,\beta=1/3$)]{\includegraphics[width=0.315\textwidth, valign=t]{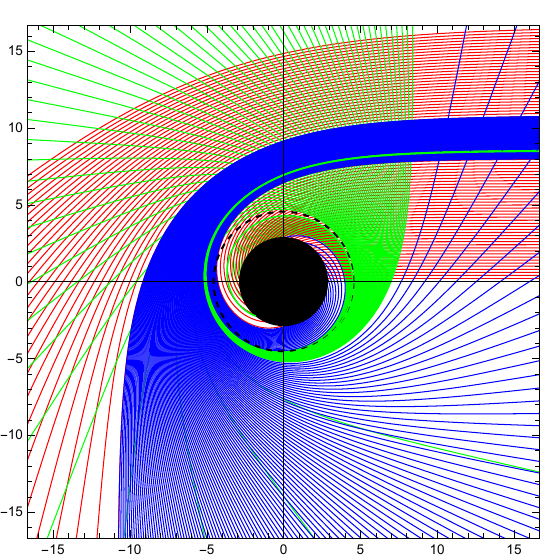}}\hfill
\subfloat[\textbf{Type III} ($n=1,\beta=3/4$)]{\includegraphics[width=0.315\textwidth, valign=t]{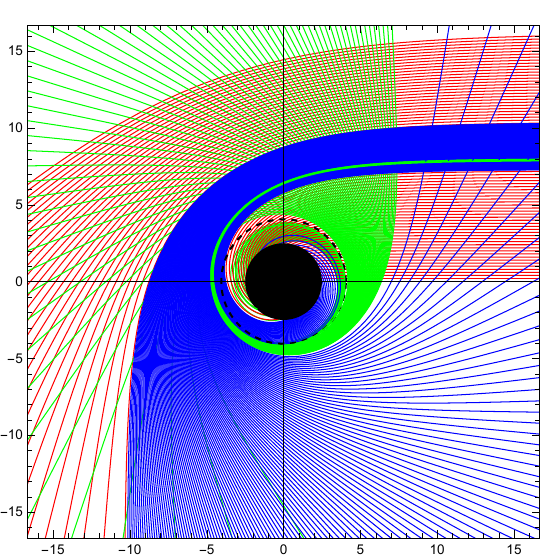}}

\vspace{-13pt}

\subfloat[\textbf{Type I} ($n=2,\beta=1/6$)]{\includegraphics[width=0.315\textwidth, valign=t]{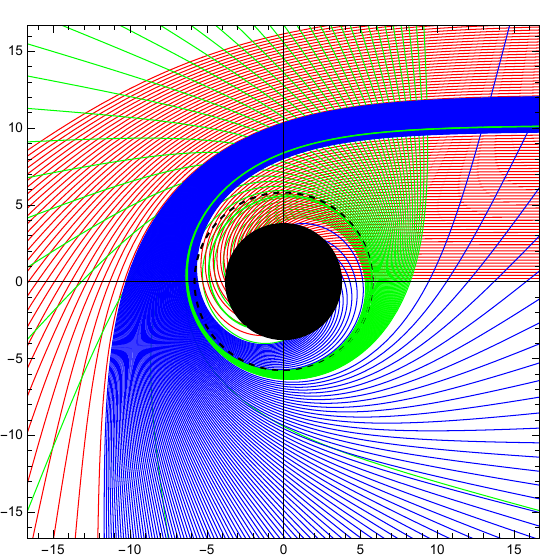}}\hfill
\subfloat[\textbf{Type II} ($n=2,\beta=2/3$)]{\includegraphics[width=0.315\textwidth, valign=t]{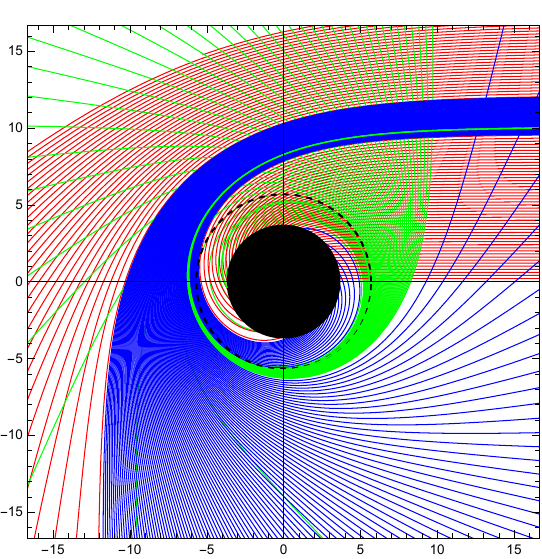}}\hfill
\subfloat[\textbf{Type III} ($n=2,\beta=3/4$)]{\includegraphics[width=0.315\textwidth, valign=t]{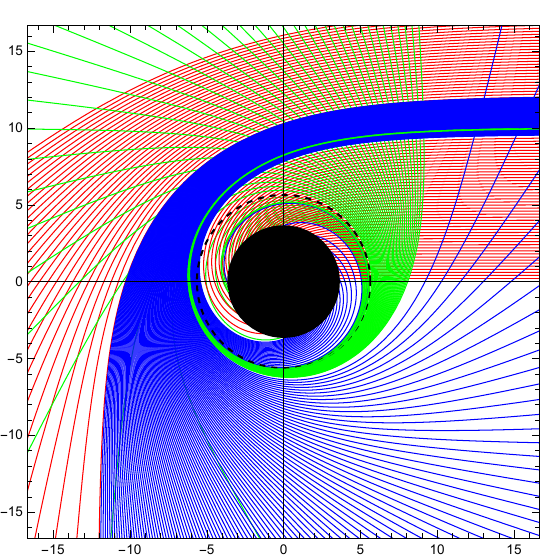}}
\vspace{-6pt}
\caption{ Upper panels: total number of orbits $N(b)$ for different $n$. The upper and lower black dashed lines mark the critical values $N = 5/4$ and $N = 3/4$, respectively. Middle and lower panels: Corresponding photon trajectories in polar coordinates $(b,\phi)$; red, blue and green curves denote direct emission, lensed ring and photon ring, respectively. The black dashed circle indicates the photon sphere ($b=b_{c}$) and the central black disk the event horizon. We set $M=2$, $\alpha=0.7$.}
\vspace{-15pt}
\label{Fig:7}
\end{figure}
Fig.~\ref{Fig:7} displays the number of orbits $N(b)$ and the associated photon trajectories for different values of $n$. The curve segments with $N > 5/4$, $3/4 < N < 5/4$, and $N < 3/4$ correspond to the photon ring, lensed ring, and direct emission, respectively. As $b$ increases, $N$ first rises, diverges as $b$ approaches the critical impact parameter $b_c$, and then gradually decreases for $b > b_c$. The corresponding photon trajectories with $b < b_c$ fall into the BH, while those with $b > b_c$ escape to infinity after deflection.

Fig.~\ref{Fig:7} reveals systematic dependencies on the parameter $n$. For a fixed $n$, the critical impact parameter $b_c$ is largest for \textbf{Type~I} and smallest for \textbf{Type~III}, resulting in the narrowest photon and lensed rings for \textbf{Type~I} and broadest for \textbf{Type~III}. As $n$ increases, $b_c$ increases for all three types while the ring widths decrease. Moreover, the distinctions in $b_c$ among the three types diminish with increasing $n$, implying that larger $n$ makes the BHs harder to distinguish based on the size of photon ring.
\begin{figure}[H]
\centering
\vspace{-18pt}
\subfloat[$\alpha=0.3$]{\includegraphics[width=0.315\textwidth, valign=t]{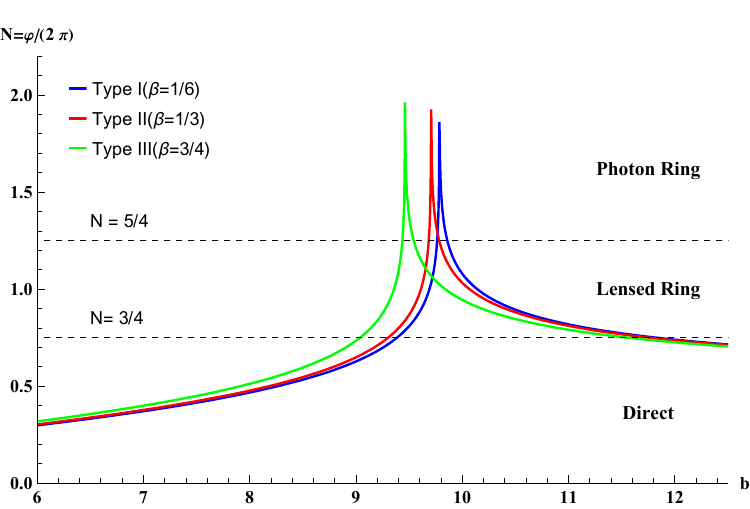}}\hfill
\subfloat[$\alpha=0.5$]{\includegraphics[width=0.315\textwidth, valign=t]{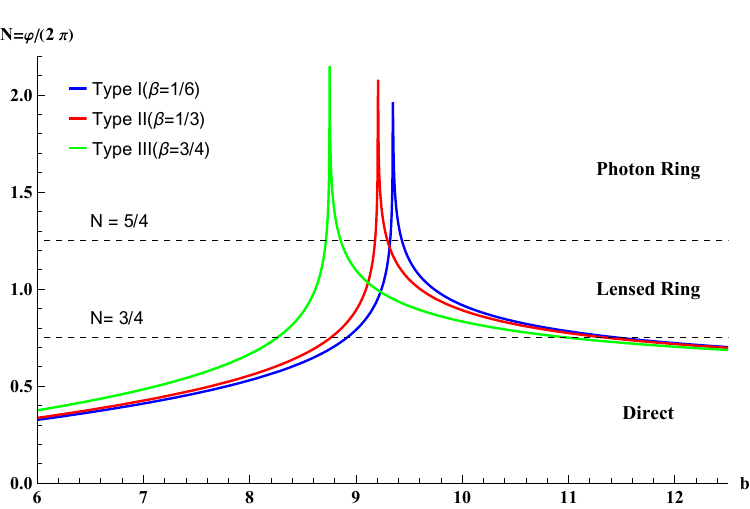}}\hfill
\subfloat[$\alpha=0.7$]{\includegraphics[width=0.315\textwidth, valign=t]{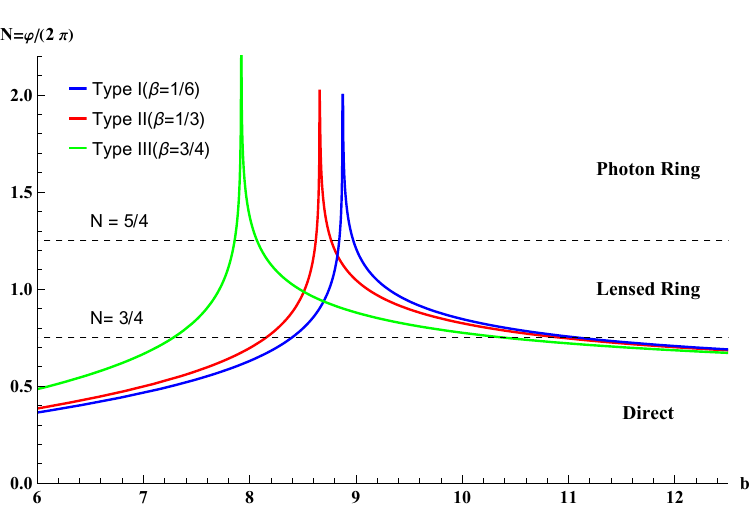}}

\vspace{-15pt}

\subfloat[\textbf{Type I} ($\alpha=0.3,\beta=1/6$)]{\includegraphics[width=0.315\textwidth, valign=t]{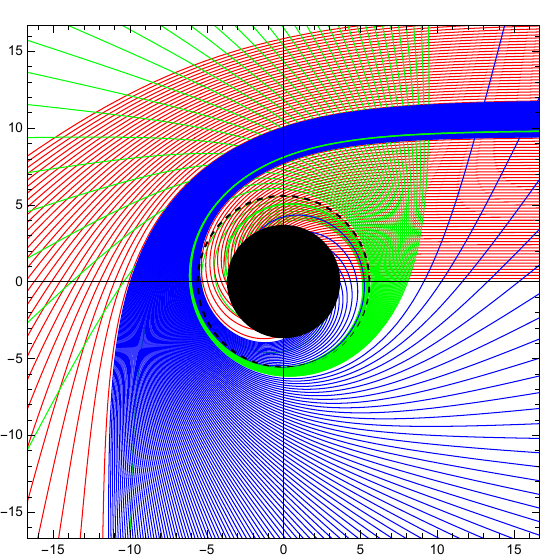}}\hfill
\subfloat[\textbf{Type II} ($\alpha=0.3,\beta=1/3$)]{\includegraphics[width=0.315\textwidth, valign=t]{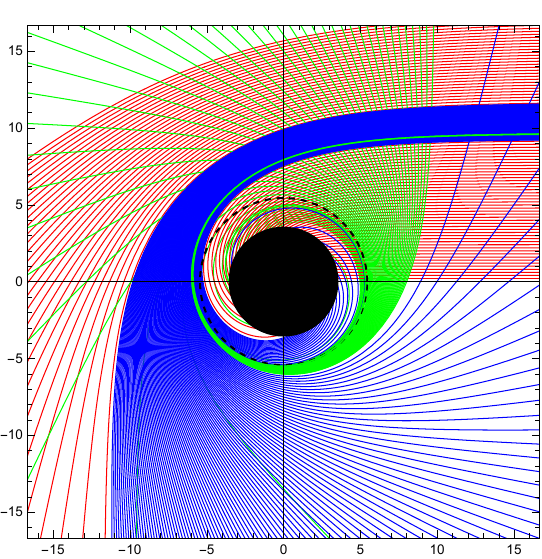}}\hfill
\subfloat[\textbf{Type III} ($\alpha=0.3,\beta=3/4$)]{\includegraphics[width=0.315\textwidth, valign=t]{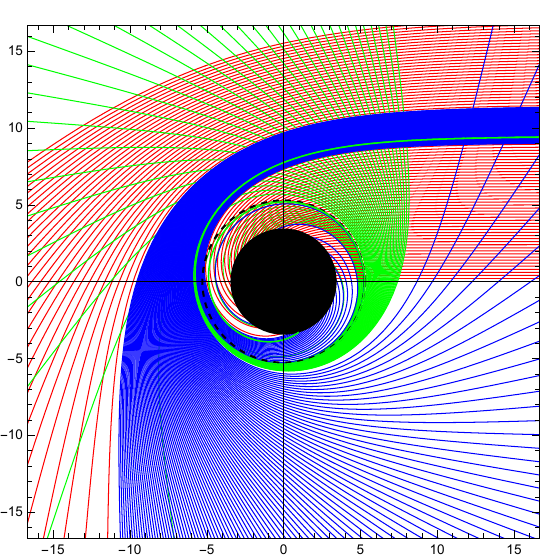}}

\vspace{-15pt}

\subfloat[\textbf{Type I} ($\alpha=0.7,\beta=1/6$)]{\includegraphics[width=0.315\textwidth, valign=t]{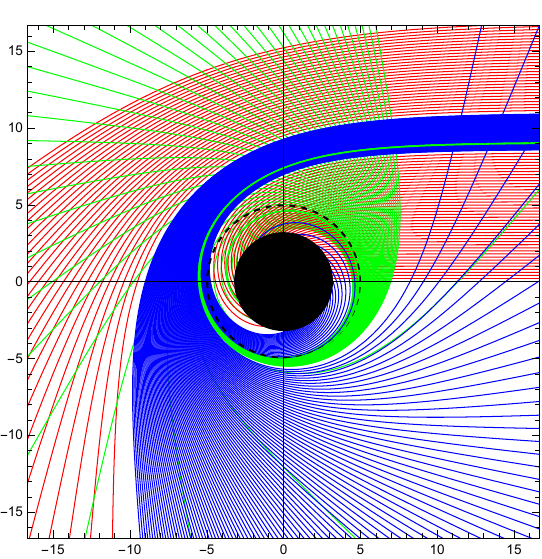}}\hfill
\subfloat[\textbf{Type II} ($\alpha=0.7,\beta=1/3$)]{\includegraphics[width=0.315\textwidth, valign=t]{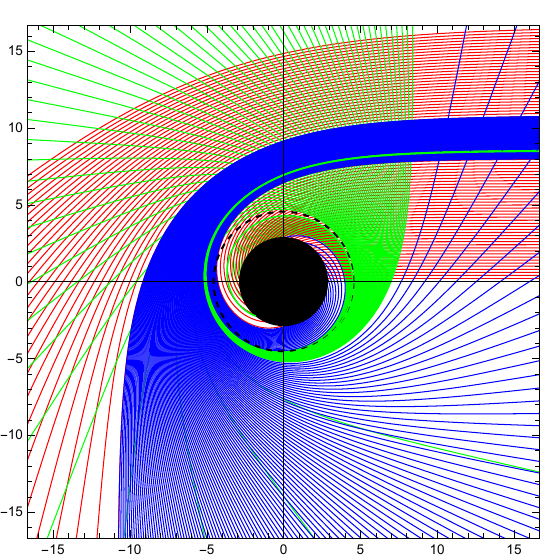}}\hfill
\subfloat[\textbf{Type III} ($\alpha=0.7,\beta=3/4$)]{\includegraphics[width=0.315\textwidth, valign=t]{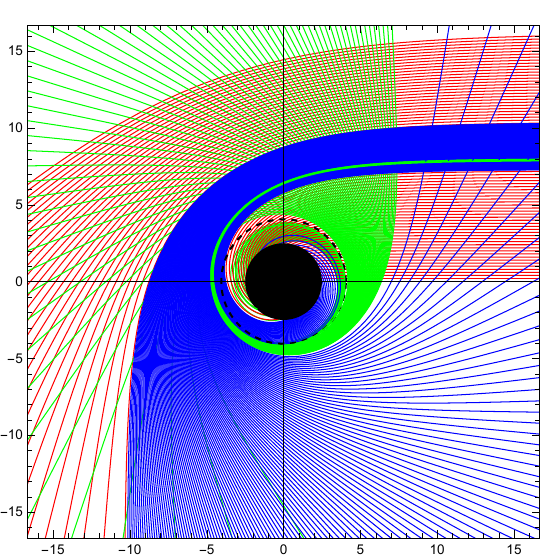}}
\vspace{-8pt}
\caption{The total number of orbits $N(b)$ and the corresponding photon trajectories in polar coordinates $(b, \phi)$ for different values of $\alpha$ ($M=2$, $n=1$).}
\vspace{-22pt}
\label{Fig:8}
\end{figure}

Fig.~\ref{Fig:8} presents analogous results for variations in $\alpha$. As $\alpha$ increases, the critical impact parameter $b_c$ decreases for all three BH types, while the widths of both the photon ring and the lensed ring increase. Notably, the differences in $b_c$ among the three types become progressively more pronounced with larger $\alpha$. This indicates that a higher value of $\alpha$ enhances the distinguishability of the three BH types based on the size of their photon rings. The clear ordering in $b_c$ $(\textbf{Type~I}>\textbf{Type~II}>\textbf{Type~III})$ and the opposite ordering in ring width provide a direct kinematic prediction for the expected shadow sizes and brightness distributions under accretion.

\section{Optical appearance and shadow formation}
\label{sec3}
Astrophysical BHs are typically surrounded by substantial accretion matter. To investigate how such environments affect observable shadows, simplified spherical accretion models are widely employed. In this section we compute the shadows and optical appearance of the three types of nonsingular BHs introduced above, assuming they are illuminated by an optically thin, geometrically spherical accretion flow. Our goal is to determine how the spacetime parameters $n$ and $\alpha$ shape the observed features and whether those features can be used to distinguish the three BH types.

\subsection{Static spherical accretion model}
\label{sec3-1}

We first consider a static spherical accretion model. For a distant observer, the observed specific intensity $I_{obs}$ (in units erg  s$^{-1}$cm$^{-2}$str$^{-1}$Hz$^{-1}$) is obtained by integrating the emissivity along the photon path $\gamma$ \cite{83,84}
\begin{equation}
\vspace{-5pt}
    I_{obs}(v_{obs})=\int_{\gamma }^{} g^{3} j(v_{e} ) \mathrm{d}l_{prop}.
    \label{eq:19}
\end{equation}
Here, $g=v_{obs} /v_{e} =\sqrt{f(r)} $ is the redshift factor, $v_{e}$ and $v_{obs}$ are the emitted and observed photon frequencies, respectively. Assuming that the radial distribution of frequency radiation is $1/r^2$, then the emissivity per unit volume $j(v_e) \propto \delta(v_e - v)/r^2$ \cite{84}, where $v$ is the rest-frame frequency and $\delta$ is the delta function. The $\mathrm{d}l_{prop}$ refers to the infinitesimal proper length, which is expressed as
\begin{equation}
   \mathrm{d}l_{prop}=\sqrt{f(r)^{-1}\mathrm{d}r^{2}+ r^{2} \mathrm{d}\phi^{2} } =\sqrt{f(r)^{-1}+r^{2}(\frac{\mathrm{d}\phi }{\mathrm{d}r} )^2 }\mathrm{d}r.
    \label{eq:20}
\end{equation}
From Eqs.~(\ref{eq:19}) and ~(\ref{eq:20}), the observed specific intensity becomes
\begin{equation}
   I_{obs} =\int_{\gamma }^{} \frac{f(r)^{\frac{3}{2} }}{r^{2} }  \sqrt{f(r)^{-1}+r^{2}(\frac{\mathrm{d} \phi  }{\mathrm{d}r} )^{2}}\mathrm{d}r.
   \label{eq:21}
\end{equation}
Using Eq.~(\ref{eq:21}), we investigate the intensity and two-dimensional shadow images, with particular attention to how variations in the spacetime parameters $n$ and $\alpha$ affect the results for the three BH types.
\begin{figure}[H]
\centering
\vspace{-10pt}
\subfloat[$n=1$]{\includegraphics[width=0.315\linewidth, valign=t]{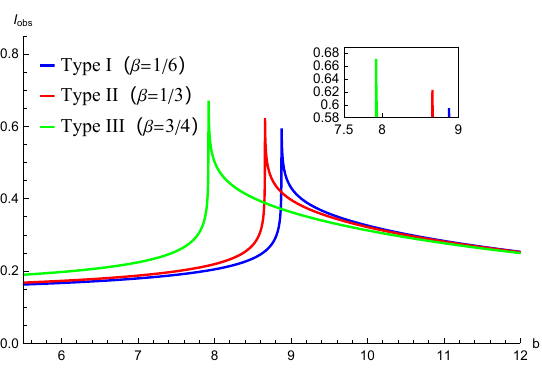}}\hfill
\subfloat[$n=2$]{\includegraphics[width=0.315\linewidth, valign=t]{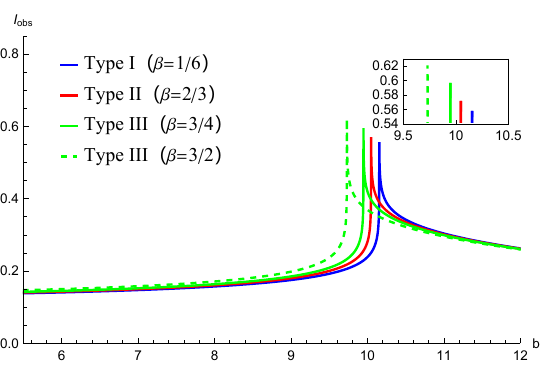}}\hfill
\subfloat[$n=3$]{\includegraphics[width=0.315\linewidth, valign=t]{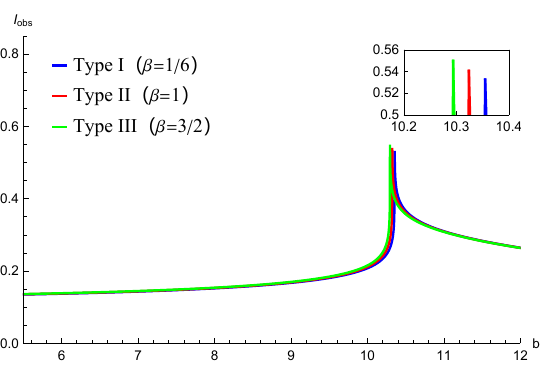}}

\vspace{-12pt}

\subfloat[\textbf{Type I} ($n=1,\beta=1/6$)]{\includegraphics[width=0.315\linewidth, valign=t]{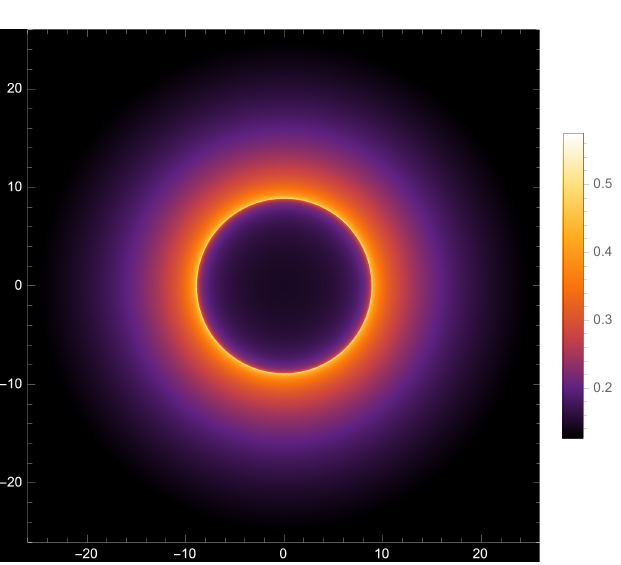}}\hfill
\subfloat[\textbf{Type II} ($n=1,\beta=1/3$)]{\includegraphics[width=0.315\linewidth, valign=t]{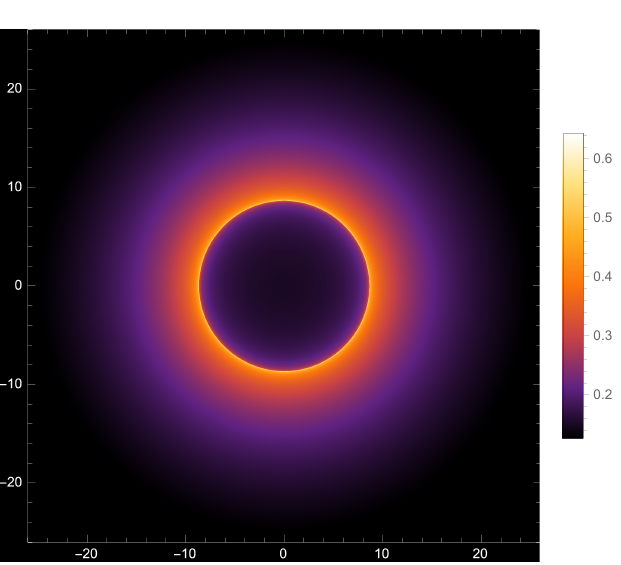}}\hfill
\subfloat[\textbf{Type III} ($n=1,\beta=3/4$)]{\includegraphics[width=0.315\linewidth, valign=t]{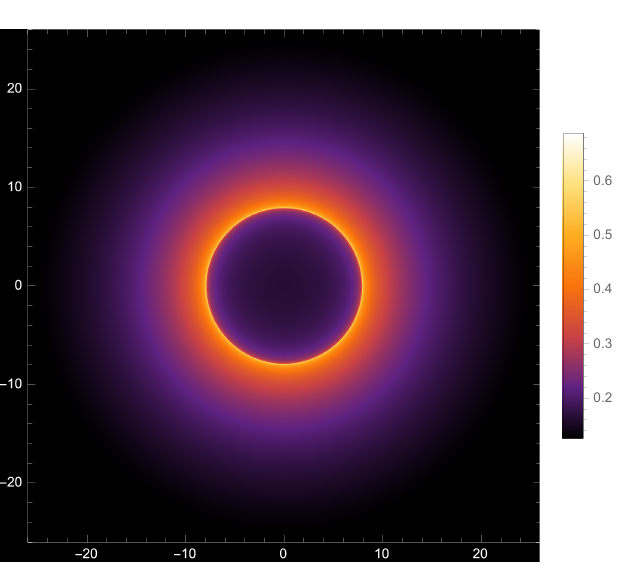}}

\vspace{-12pt}

\subfloat[\textbf{Type I} ($n=2,\beta=1/6$)]{\includegraphics[width=0.315\linewidth, valign=t]{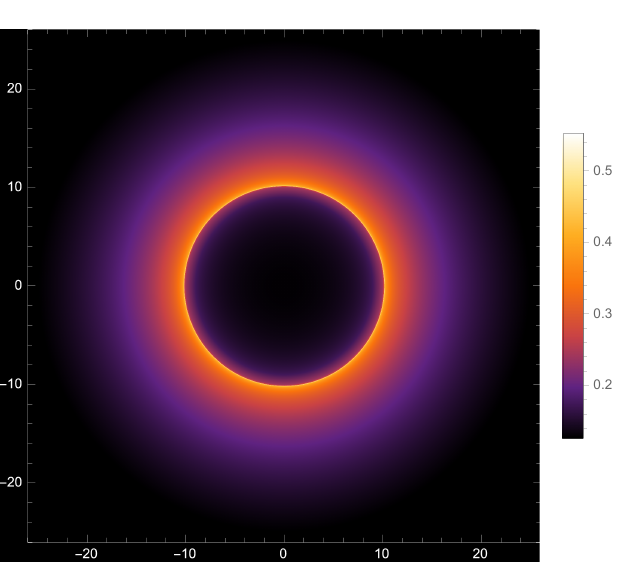}}\hfill
\subfloat[\textbf{Type II} ($n=2,\beta=2/3$)]{\includegraphics[width=0.315\linewidth, valign=t]{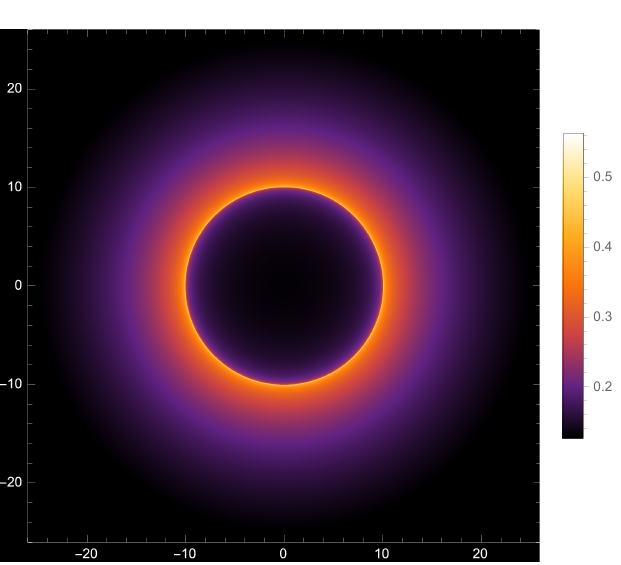}}\hfill
\subfloat[\textbf{Type III} ($n=2,\beta=3/4$)]{\includegraphics[width=0.315\linewidth, valign=t]{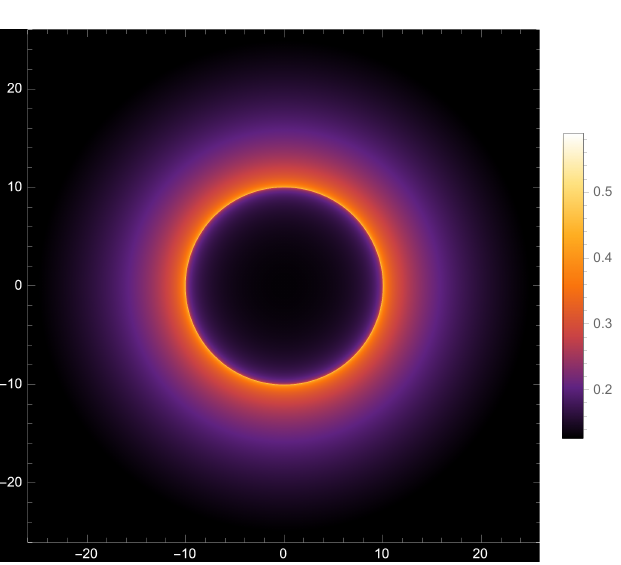}}
\vspace{0pt}
\caption{The observed specific intensity and shadow images under static spherical accretion for different values of $n$ ($M=2$, $\alpha=0.7$). Upper panels: The observed specific intensity. Middle and lower panels: Corresponding two-dimensional shadow images for the three types. The bright ring marks the photon ring. }
\vspace{-10pt}
\label{Fig:9}
\end{figure}
Fig.~\ref{Fig:9} presents the observed specific intensity and shadow images of three BH types under different values of $n$. The observed specific intensity $I_{obs}$ (upper panels) rises gradually with $b$, peaks near the critical impact parameter $b_c$, and then decreases. For a fixed $n$, the peak intensity is weakest for \textbf{Type~I} and strongest for \textbf{Type~III}, while the corresponding $b_c$ follows the opposite trend: largest for \textbf{Type~I} and smallest for \textbf{Type~III}. The two-dimensional shadow images (middle and lower panels) clearly display a bright photon ring surrounding the central dark region, where the luminosity is highest.

\begin{figure}[H]
\centering
\vspace{-16pt}
\subfloat[$\alpha=0.3$]{\includegraphics[width=0.315\linewidth, valign=t]{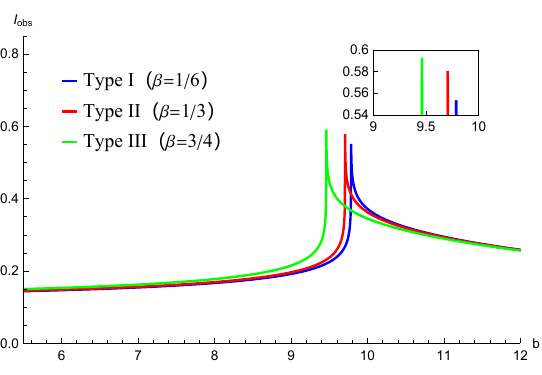}}\hfill
\subfloat[$\alpha=0.5$]{\includegraphics[width=0.315\linewidth, valign=t]{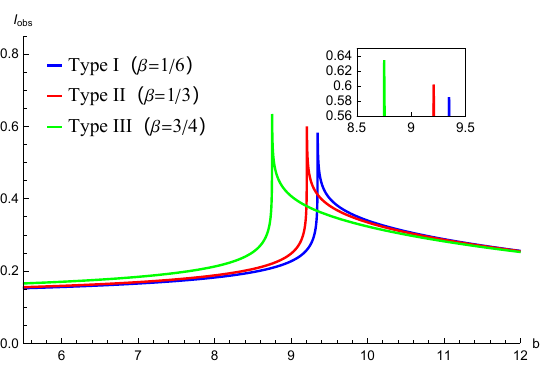}}\hfill
\subfloat[$\alpha=0.7$]{\includegraphics[width=0.315\linewidth, valign=t]{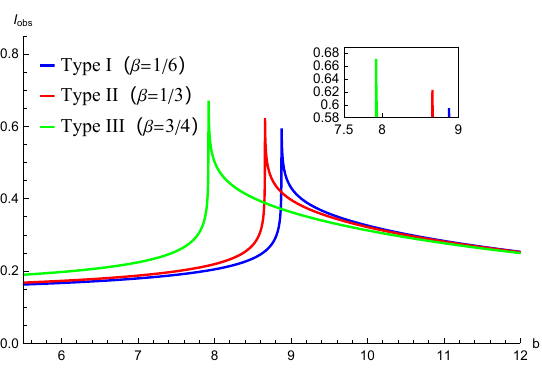}}

\vspace{-15pt}

\subfloat[\textbf{Type I} ($\alpha=0.3,\beta=1/6$)]{\includegraphics[width=0.315\linewidth, valign=t]{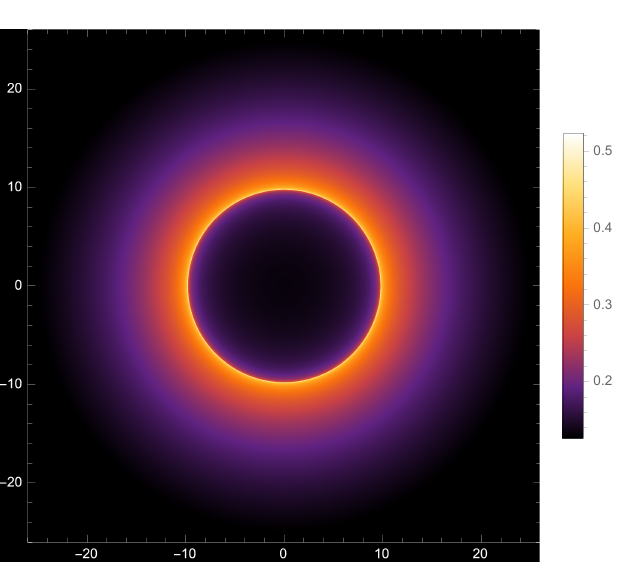}}\hfill
\subfloat[\textbf{Type II} ($\alpha=0.3,\beta=1/3$)]{\includegraphics[width=0.315\linewidth, valign=t]{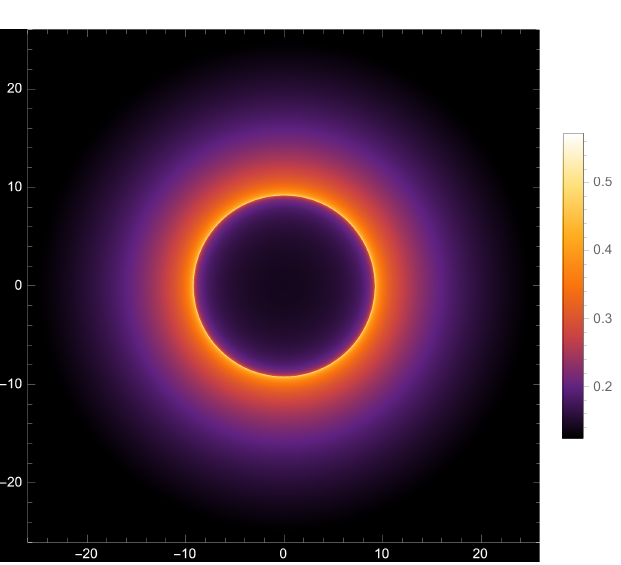}}\hfill
\subfloat[\textbf{Type III} ($\alpha=0.3,\beta=3/4$)]{\includegraphics[width=0.315\linewidth, valign=t]{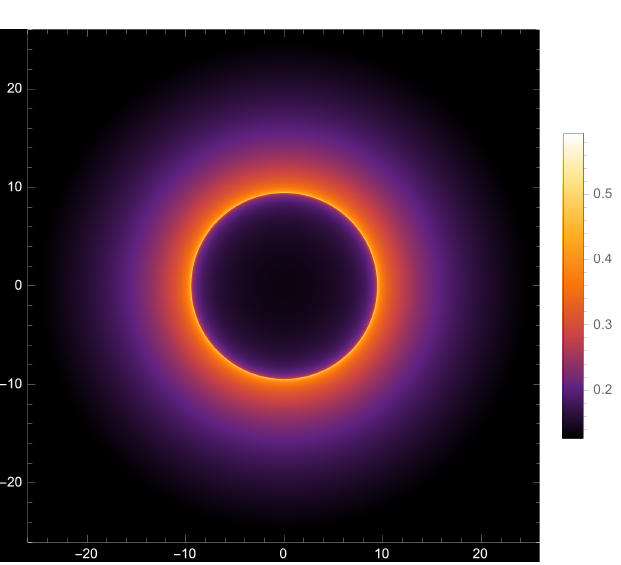}}

\vspace{-15pt}

\subfloat[\textbf{Type I} ($\alpha=0.7,\beta=1/6$)]{\includegraphics[width=0.315\linewidth, valign=t]{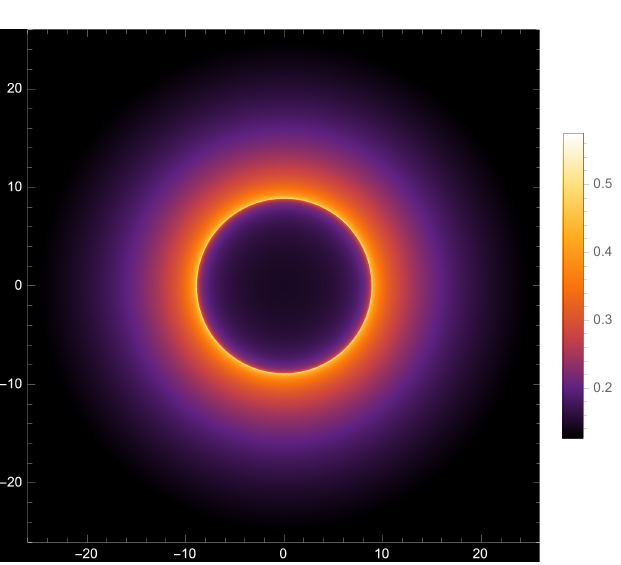}}\hfill
\subfloat[\textbf{Type II} ($\alpha=0.7,\beta=1/3$)]{\includegraphics[width=0.315\linewidth, valign=t]{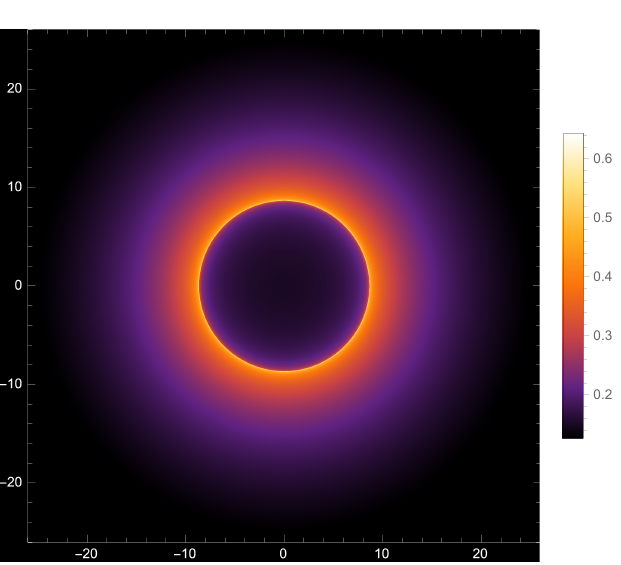}}\hfill
\subfloat[\textbf{Type III} ($\alpha=0.7,\beta=3/4$)]{\includegraphics[width=0.315\linewidth, valign=t]{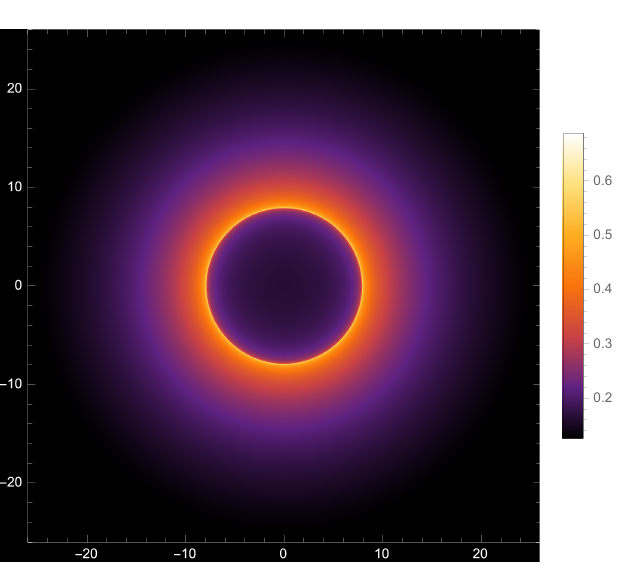}}
\vspace{-8pt}
\caption{The observed specific intensity and shadow images under static spherical accretion for different values of $\alpha$ ($M=2$, $n=1$).}
\vspace{-20pt}
\label{Fig:10}
\end{figure}
The comparison across Fig.~\ref{Fig:9} reveals the influence of the parameter $n$ on the optical appearance. As $n$ increases, the observed intensity gradually decreases for all three BH types, while the critical impact parameter $b_c$ increases. At the same time, the differences among three types, both in observed intensity and photon ring's characteristics, become more pronounced with smaller $n$. This indicates that the three BH types can be more easily distinguished based on their optical appearances when $n$ is smaller. Furthermore, the comparison of the optical appearances under different values of $n$ further reveals that the \textbf{Type III} BHs undergo the most noticeable changes in both observed intensity and photon ring radius as $n$ varies. This shows that the optical characteristics of \textbf{Type III} BHs are more sensitive to the parameter $n$ than those of the other two types.

The influence of the parameter $\alpha$ is shown in Fig.~\ref{Fig:10}. As $\alpha$ increases, the observed intensity rises for all three BH types. In contrast, the photon ring radius decreases while its luminosity increases. Consequently, the distinctions in optical appearance among the three BH types become increasingly evident for larger $\alpha$, making them more readily distinguishable. Moreover, the \textbf{Type~III} BHs exhibit the most pronounced variation in both observed intensity and photon ring radius as the parameter $\alpha$ changes. This demonstrates that the optical characteristics of \textbf{Type~III} BHs are subject to the greatest influence from $\alpha$ among the three BH types.
\subsection{Infalling spherical accretion model}
\label{sec3-2}

To better approximate realistic astrophysical conditions, we now consider an infalling spherical accretion model. In this dynamical model, the radiating gas moves radially toward the BH. While the fundamental expression for the observed intensity Eq.~(\ref{eq:19}) remains valid, the redshift factor must be modified to
\begin{equation}
g=\frac{k_{\zeta } u^{\zeta }_{\text{obs}}}{k_{\eta} u^{\eta }_{e}} .
\label{eq:22}
\end{equation}
Here, $k^{\mu} \equiv \dot{x}^{\mu}$ denotes the four-momentum of the photon. According to the null geodesic condition, $k_{\mu}k^{\mu}=0$. Using Eq.~(\ref{eq:11}), $k_{\mu }$ can be written as
\begin{equation}
k_{t}=\frac{1}{b}, \quad
k_r=\pm \frac{\sqrt{r^2-b^2 f(r)}}{b\,r\,f(r)}.
\label{eq:23}
\end{equation}
The ``$\pm$'' sign indicates that the photon can either approach or escape from the BH. In addition, $u^{\mu}_{\text{obs}} \equiv (1,0,0,0)$ represents the four-velocity of a static distant observer and $u^{\mu}_{\text{e}}$ corresponds to the four-velocity of the infalling accretion, which has a form
\begin{equation}
u^{t}_{e}=f(r)^{-1}=\left(1-\frac{2M \mathrm{e}^{-\frac{\alpha M^{\beta}}{r^{n}}}}{r}\right)^{-1},
\quad
u^{r}_{e}=-\sqrt{1-f(r)}=-\sqrt{\frac{2M \mathrm{e}^{-\frac{\alpha M^{\beta}}{r^{n}}}}{r}},
\quad
u^{\theta}_{e}=u^{\varphi}_{e}=0 .
\label{eq:24}
\end{equation}
Then, the redshift factor for the infalling spherical accretion can be expressed as
\begin{equation}
g=\frac{k_{t}}{k_{t} u^{t}_{e}+k_{r} u^{r}_{e} } .
\label{eq:25}
\end{equation}
The infinitesimal proper length along the photon trajectory is rewritten as \cite{84}
\begin{equation}
\mathrm{d}l_{\text{prop}}
= k_{\eta}u^{\eta}_{e}\,\mathrm{d}\lambda
= \frac{k_{t}}{g |k_{r}|} \,\mathrm{d}r .
\label{eq:26}
\end{equation}
Therefore, the observed specific intensity for the infalling spherical accretion is given by
\begin{equation}
I_{\text{obs}}
= \int_{\gamma} \frac{g^{3} k_{t} }{r^{2} |k_{r}|  } \,\mathrm{d}r .
\label{eq:27}
\end{equation}

\begin{figure}[H]
\centering
\vspace{0pt}
\subfloat[$n=1$]{\includegraphics[width=0.325\linewidth]{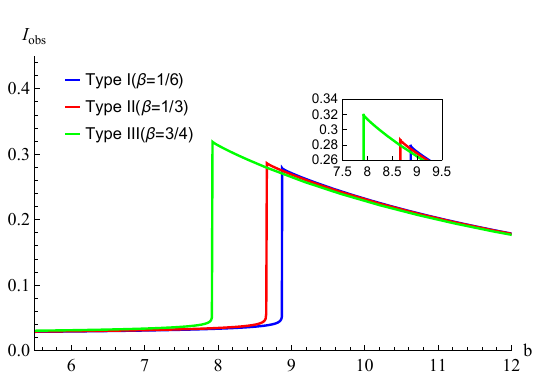}}\hfill
\subfloat[$n=2$]{\includegraphics[width=0.325\linewidth]{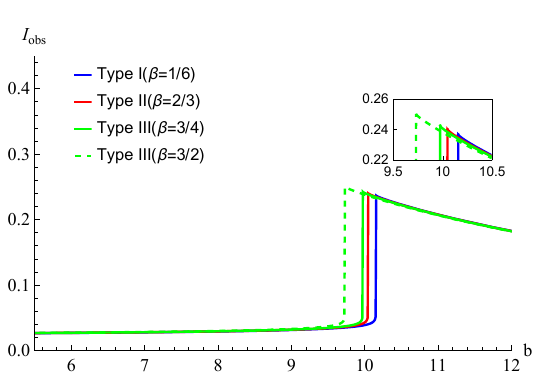}}\hfill
\subfloat[$n=3$]{\includegraphics[width=0.325\linewidth]{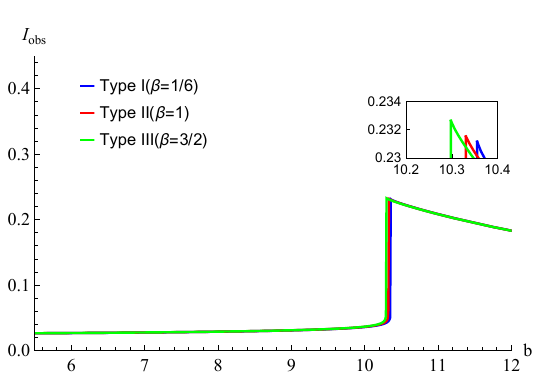}}

\vspace{-12pt}

\subfloat[\textbf{Type I} ($n=1,\beta=1/6$)]{\includegraphics[width=0.325\linewidth]{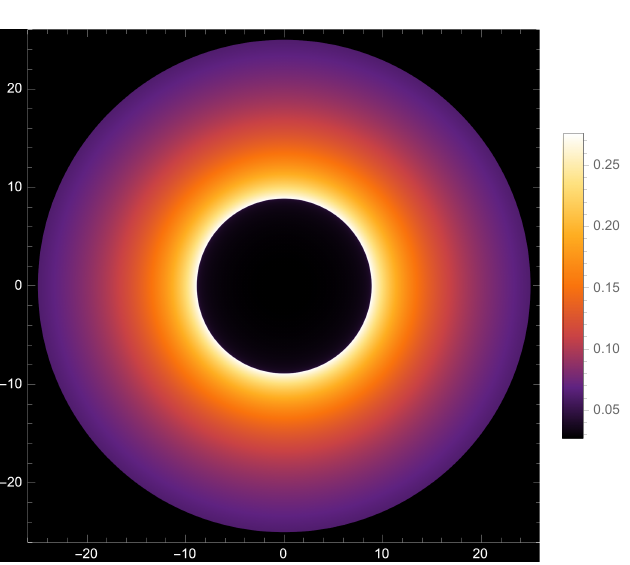}}\hfill
\subfloat[\textbf{Type II} ($n=1,\beta=1/3$)]{\includegraphics[width=0.325\linewidth]{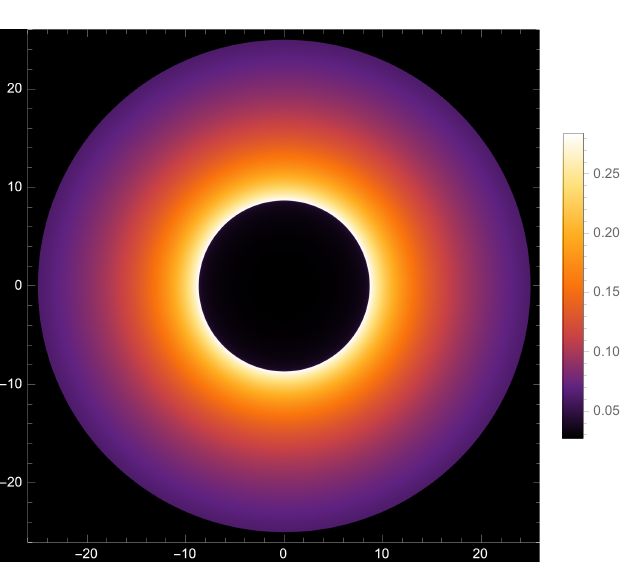}}\hfill
\subfloat[\textbf{Type III} ($n=1,\beta=3/4$)]{\includegraphics[width=0.325\linewidth]{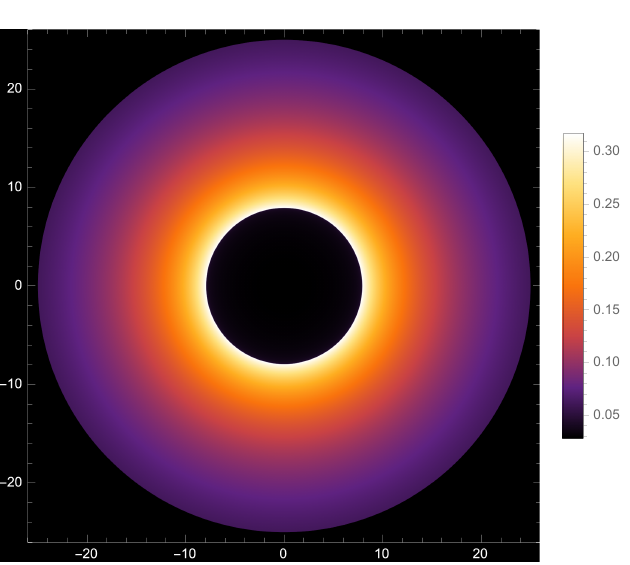}}

\vspace{-12pt}

\subfloat[\textbf{Type I} ($n=2,\beta=1/6$)]{\includegraphics[width=0.325\linewidth]{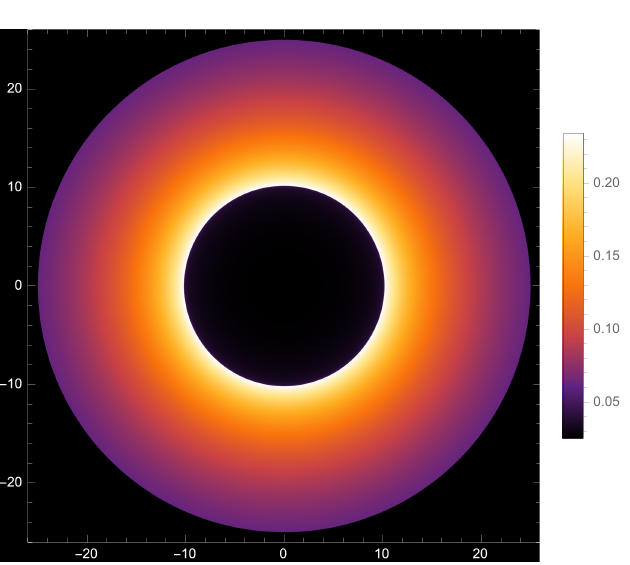}}\hfill
\subfloat[\textbf{Type II} ($n=2,\beta=2/3$)]{\includegraphics[width=0.325\linewidth]{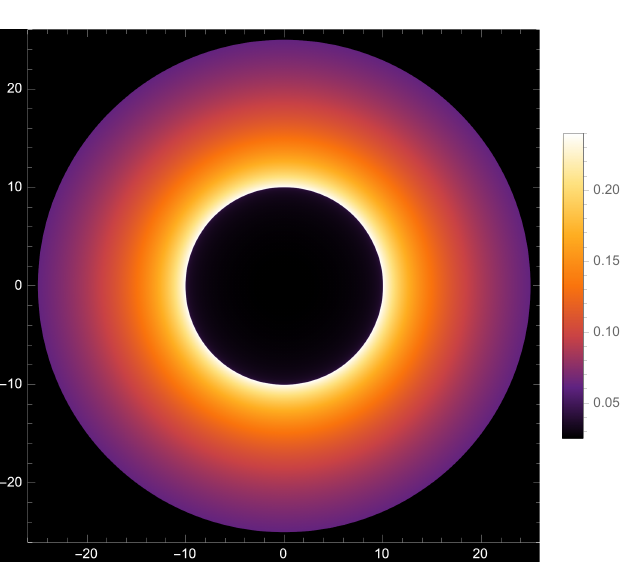}}\hfill
\subfloat[\textbf{Type III} ($n=2,\beta=3/4$)]{\includegraphics[width=0.325\linewidth]{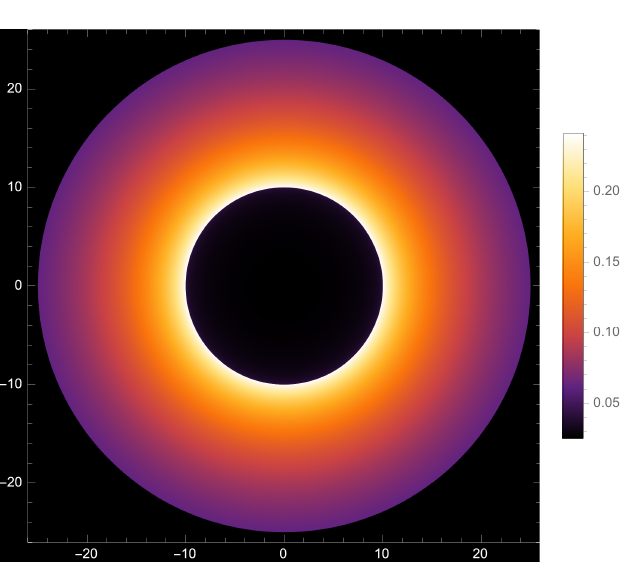}}
\vspace{-5pt}
\caption{Observed specific intensity and shadow images for the infalling spherical accretion with different $n$ ($M=2$, $\alpha=0.7$).}
\vspace{-10pt}
\label{Fig:11}
\end{figure}
\begin{figure}[H]
\centering
\vspace{-10pt}
\subfloat[$\alpha=0.3$]{\includegraphics[width=0.325\linewidth]{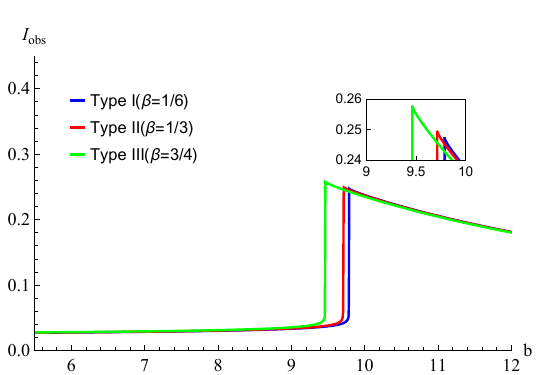}}\hfill
\subfloat[$\alpha=0.5$]{\includegraphics[width=0.325\linewidth]{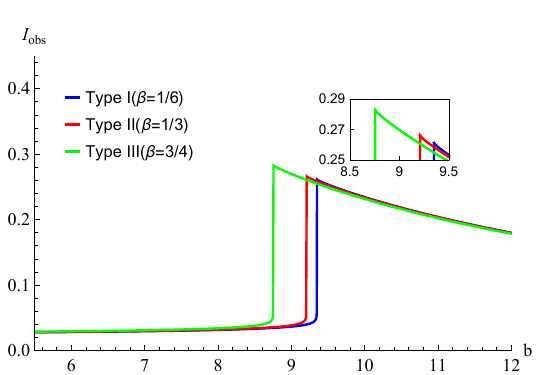}}\hfill
\subfloat[$\alpha=0.7$]{\includegraphics[width=0.325\linewidth]{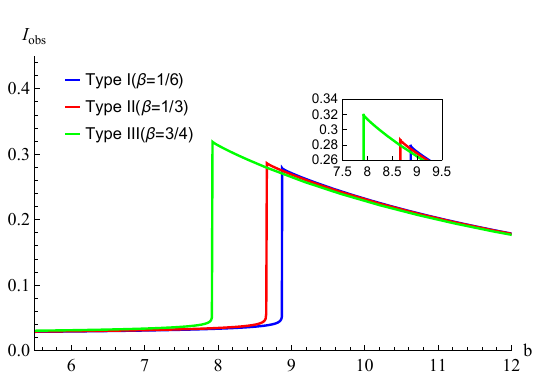}}

\vspace{-12pt}

\subfloat[\textbf{Type I} ($\alpha=0.3,\beta=1/6$)]{\includegraphics[width=0.325\linewidth]{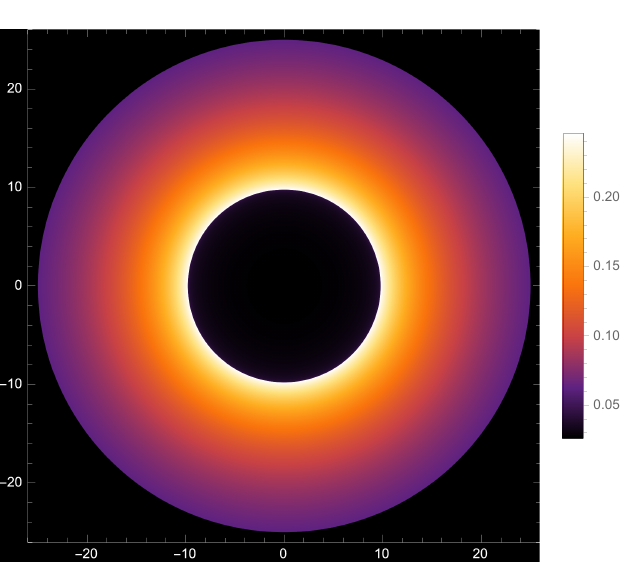}}\hfill
\subfloat[\textbf{Type II} ($\alpha=0.3,\beta=1/3$)]{\includegraphics[width=0.325\linewidth]{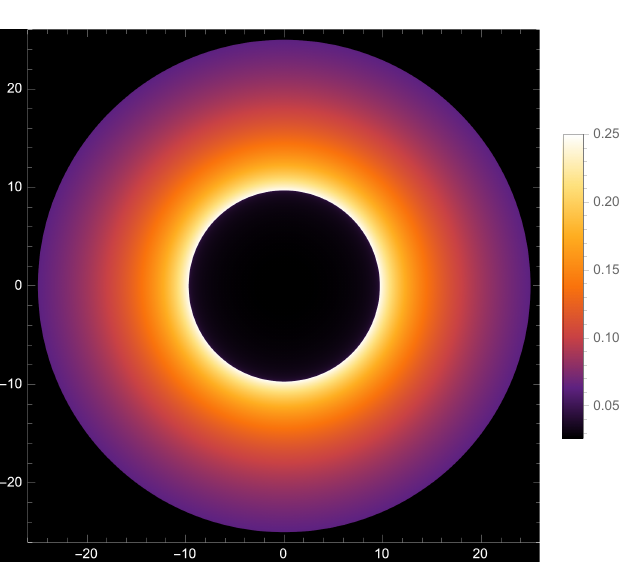}}\hfill
\subfloat[\textbf{Type III} ($\alpha=0.3,\beta=3/4$)]{\includegraphics[width=0.325\linewidth]{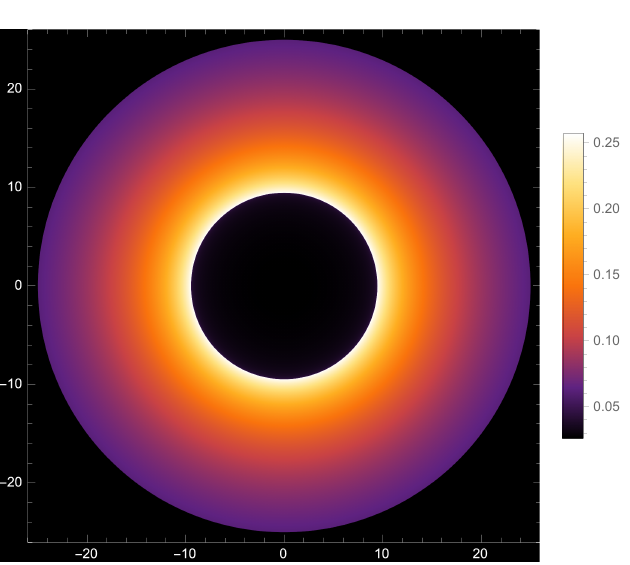}}

\vspace{-12pt}

\subfloat[\textbf{Type I} ($\alpha=0.7,\beta=1/6$)]{\includegraphics[width=0.325\linewidth]{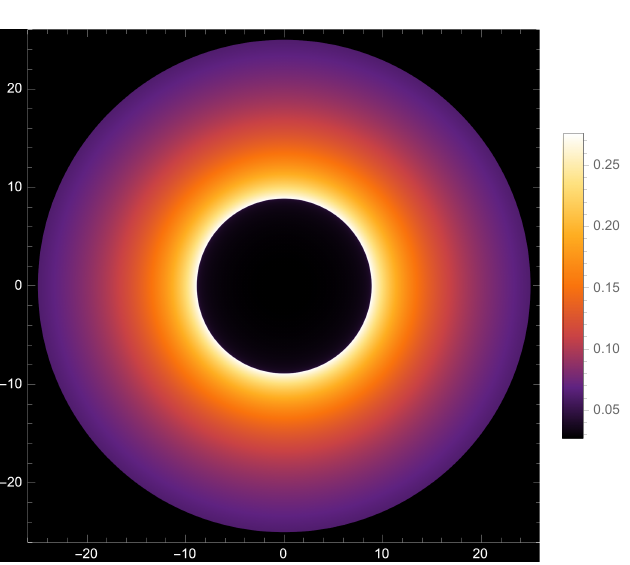}}\hfill
\subfloat[\textbf{Type II} ($\alpha=0.7,\beta=1/3$)]{\includegraphics[width=0.325\linewidth]{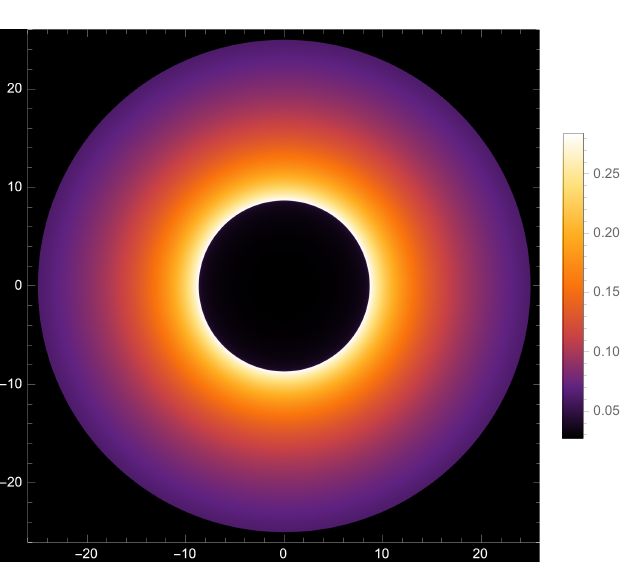}}\hfill
\subfloat[\textbf{Type III} ($\alpha=0.7,\beta=3/4$)]{\includegraphics[width=0.325\linewidth]{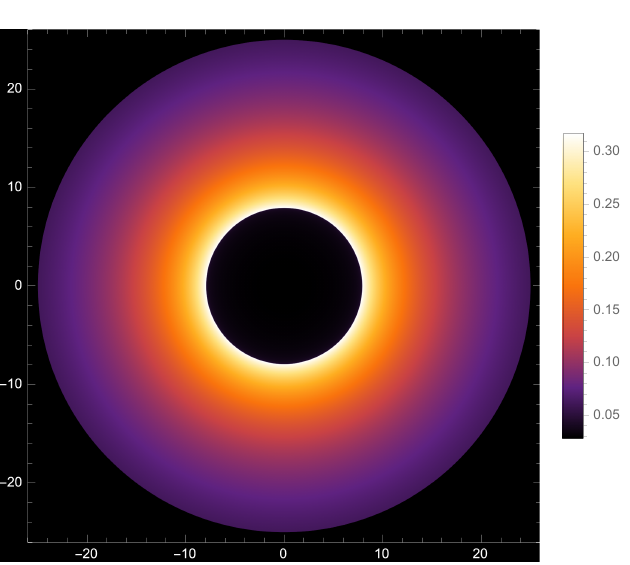}}
\vspace{-5pt}
\caption{Observed specific intensity and shadow images for the infalling spherical accretion with different $\alpha$ ($M=2$, $n=1$).}
\vspace{-10pt}
\label{Fig:12}
\end{figure}

Figs.~\ref{Fig:11} and \ref{Fig:12} present the results for the infalling spherical accretion model. While the overall intensity profiles are qualitatively similar to those obtained under the static spherical accretion model (see Figs.~\ref{Fig:9} and \ref{Fig:10}), a notable distinction is the extremely sharp rise in intensity prior to the peak, followed by a more gradual decline. The two-dimensional shadow images exhibit features similar to those of the static model. Consistent trends are observed: the observed intensity decreases with increasing $n$ but increases with $\alpha$, and \textbf{Type~III} BHs display the highest intensity and the smallest photon ring radius for fixed parameters. As $n$ increases or $\alpha$ decreases, the optical differences among the three BH types become less pronounced, rendering them increasingly difficult to distinguish. Notably, \textbf{Type~III} BHs exhibit the strongest dependence on the spacetime geometry, leading to the most pronounced variations in their optical appearance.
The pronounced sensitivity of \textbf{Type III} BHs observed in both accretion models stems from their defining geometric property: $K_{\text{max}} \propto M^{-(6\beta/n - 2)}$ with $\beta > n/3$. This inverse mass-curvature relation implies that the underlying regularization mechanism becomes more effective at larger masses, thereby producing a flatter core while simultaneously creating a steeper gradient in the external spacetime. As a result, the photon sphere in \textbf{Type III} geometries resides in a more sharply peaked region of the effective potential, where small variations in parameters $(n, \alpha)$ induce proportionally larger changes in $r_c$ and $b_c$ than in \textbf{Type I} and \textbf{Type II}. Mathematically, this is reflected in the ordering of derivatives $\partial b_c/\partial \alpha|_{\mathrm{III}} > \partial b_c/\partial \alpha|_{\mathrm{II}} > \partial b_c/\partial \alpha|_{\mathrm{I}}$, which is consistent with the intensity profiles shown in Figs.~\ref{Fig:10} and \ref{Fig:12}.

\section{Conclusions and discussions}
\label{sec4}
In this work, we have established a direct connection between the intrinsic curvature structure of nonsingular BHs with a Minkowski core and their observable optical features. By classifying these spacetimes into three types based on the mass-scaling of the maximum Kretschmann scalar $K_{\text{max}}$, and subsequently analyzing their photon dynamics and shadow formation under spherical accretion, we demonstrate that subtle differences in interior geometry imprint clear, discernible signatures on the BH shadow.

The classification, \textbf{Type~I} ($K_{\text{max}}$ increasing with $M$), \textbf{Type~II} (mass-independent $K_{\text{max}}$), and \textbf{Type~III} ($K_{\text{max}}$ decreasing with $M$), encapsulates fundamental variations in how curvature is distributed. Our detailed parameter study reveals that $\alpha$, which encodes the deviation from the Schwarzschild metric, monotonically suppresses the central curvature, leading to a smoother geometry. In contrast, the parameter $n$ governs a non-monotonic reorganization of the spacetime: curvature is most suppressed around $n \approx 2$, with higher $K_{\text{max}}$ for both smaller and larger values.

A crucial observation is that while the core curvature peak $K_{\text{max}}$ exhibits non-monotonic behavior with $n$, all external optical features vary monotonically. This highlights that the mapping from interior curvature to external observables is not direct or point-to-point. $K_{\text{max}}$ is a local invariant characterizing curvature intensity at the core, whereas shadow formation is governed by the global structure of the photon sphere. The parameters $n$ and $\alpha$ influence the photon sphere by altering the overall metric profile, and this global effect dominates the observational signal, overwhelming the local non-monotonic details near the core. Specifically, the peak value of the effective potential at the photon sphere, $V_{\text{eff}}(r)$, is highest for \textbf{Type~III} BHs and lowest for \textbf{Type~I} BHs. This order in the potential barrier directly correlates with their external geometry: a higher peak corresponds to a more compact photon sphere and a steeper gravitational gradient at the critical orbit, which is the defining feature of \textbf{Type~III} BHs.

These intrinsic geometric differences propagate outward and are reflected in the properties of the photon sphere, which acts as the key intermediary. The effective potential $V_{\text{eff}}(r)$, the photon sphere radius $r_c$, and the critical impact parameter $b_c$ all exhibit systematic dependencies on $n$ and $\alpha$ that differ among the three types. Consequently, the distinctions are most pronounced for smaller $n$ or larger $\alpha$. This geometric understanding explains the observed optical patterns: for fixed parameters, \textbf{Type~III} BHs with the most compact photon sphere show the highest observed intensity and the smallest shadow radius, whereas \textbf{Type~I} BHs exhibit the opposite trend. The exceptional sensitivity of \textbf{Type~III} BHs to parameter changes stems directly from their defining geometric feature: $K_{\text{max}}$ decreasing with $M$. This inverse mass curvature relation implies that their regularization mechanism becomes more effective at larger masses, dispersing the effective mass-energy away from the core and producing a flatter central geometry. To maintain this ``high-mass, low-curvature'' configuration, the external spacetime develops an exceptionally steep profile, resulting in the most compact photon sphere and the strongest curvature gradients in its vicinity. Consequently, any variation in parameters $(n, \alpha)$ is strongly amplified there, leading to the largest changes in shadow size and brightness.

The excellent agreement of our calculated shadows under both static and infalling spherical accretion models robustly confirms that the curvature-based classification maps directly onto observable differences. Most notably, \textbf{Type~III} BHs consistently display the strongest variation in both intensity and shadow size with changing parameters, making them the most responsive probes of the underlying spacetime.

Our results carry several important implications. Theoretically, they provide a concrete example of how a fundamental curvature-invariant classification translates into a set of observable shadows, establishing a potential ``inverse-problem'' framework whereby measuring shadow size and luminosity profiles could, in principle, be used to infer the interior curvature type ($K_{\text{max}}$ scaling) of an observed BH. Observationally, the pronounced differences among the three types suggest that high-resolution instruments like the EHT may be able to distinguish between such nonsingular models.

Our findings naturally point to several promising research directions. First, a systematic Bayesian parameter-estimation study using simulated EHT-like data would quantify how precisely the parameters $(n, \alpha)$ and the curvature type itself can be constrained from actual observations. Second, extending the shadow computation to more realistic, anisotropic accretion flows (e.g., magnetized, rotating disks) will test the robustness of the identified type-distinguishing features against astrophysical complexities. Third, a deeper physical interpretation could be achieved by linking the phenomenological parameters $n$, $\alpha$, and $\beta$ to specific quantum-gravity models or effective field theories. Finally, it would be fruitful to investigate whether an analogous curvature-based classification induces a similar observational taxonomy in other families of regular BHs (e.g., those with de Sitter cores), potentially revealing a universal principle linking interior geometry to shadow morphology.

In summary, this work moves beyond the question of distinguishing singular from nonsingular BHs and demonstrates that even among nonsingular BHs with the same asymptotically flat (Minkowski) interior, internal curvature differences leave an observable imprint. By bridging abstract spacetime classification with concrete observational signatures, we provide a new pathway to probe the elusive interior structure of BHs and to test quantum-gravity-inspired models with forthcoming shadow observations.

\section*{Acknowledgements}
This work is supported by Sichuan Science and Technology Program(2023NSFSC1352), and by the starting fund of China West Normal University (Nos. 20E069 and 20A013).

%
%

%
%
\end{document}